\documentclass{cernyrep} 
\usepackage{texnames}
\usepackage[T1]{fontenc}
\begin{document}
\title{Ultra-High Energy Cosmic Rays}
 
\author{M.T. Dova}

\institute{Instituto de F\'{\i}sica La Plata, Universidad Nacional de La Plata and CONICET, Argentina}

\maketitle 

\begin{abstract}
The origin of the ultra high energy cosmic rays (UHECR) with energies above $E>10^{17}$eV, is still unknown. The discovery of their sources will reveal the engines of the most energetic astrophysical accelerators in the universe. This is a written version of a series of lectures devoted to UHECR at the 2013 CERN-Latin-American School of High-Energy Physics. We present an introduction to  acceleration mechanisms of charged particles to the highest energies in astrophysical objects, their propagation from the sources to Earth, and the experimental techniques for their detection. We also discuss some of the relevant observational results from Telescope Array and Pierre Auger Observatory.  These experiments deal with particle interactions at energies orders of magnitude higher than achieved in terrestrial accelerators.  \end{abstract}
 
\section{Introduction}
 
Extreme physical systems provide the best scenario to study the fundamental physical laws. In this direction the research on ultra high energy cosmic rays is a crucial element, contributing to progress in both astrophysics and  particle physics. UHECR open a window to energy and kinematic regions previously unexplored in the study of fundamental 
interactions and continue to motivate current and future cosmic ray experiments. In this note we summarize a series of lectures given at the 7th CERN-Latin-American School of High-Energy Physics on ultra high energy cosmic rays,  
 the highest-energy particles measured on Earth with energy $E>10^{17}$eV. 
 
 UHECR are mainly protons and nuclei, accelerated in astrophysical objects. The requirements for these objects to be sources of UHECR are quite stringent, as in addition to be able to accelerate to extremely high energies, they should also have the luminosity that can account for the observed fluxes. UHECR must survive during acceleration, escape and propagation  through the intergalactic space,
losing energy in the interactions with the Infrared/optical (IR/O), Cosmic Microwave Background (CMB) or Radio Background photons.  We begin with a brief introduction to cosmic rays. Then, we introduce basic concepts of acceleration mechanisms,  and the main energy loss processes for UHECR during propagation. The opacity of the CMB to the propagation of these particles is a key issue in the search for the origin of UHECR, leading to a modification of the energy spectrum and a strong constraint on the proximity of UHECR sources. At this point we give a short description of the main experimental techniques for the detection of UHECR and discuss observational results of the cosmic ray spectrum.  UHECR are also deflected in the intergalactic and galactic magnetic fields in the propagation volume, what limits the search for correlations of the arrival direction of UHECR with possible sources  and distributions of astrophysical objects in our vicinity. Here we present studies of anisotropy at the highest energies. Next, we summarize the phenomenology of cosmic ray air showers, including the dominant electromagnetic processes driving the shower evolution. We also present the hadronic interaction models used to extrapolate results from collider data to ultrahigh energies. Finally, we describe the main observables sensitive to primary composition, the most challenging issue to understand the nature and origin of  UHECR.

\section{Cosmic Rays}

In 1912, Victor Hess  carried out a series of balloon flights taking an electroscope to measured the ionizing radiation as a function of altitude. He discovered  that the ionization rate increased by at least a factor of two at around 5 km above the Earth's surface ~\cite{bib:hess}. He received the Nobel prize in 1936 for the discovery of this ``penetrating radiation'' coming from space, later called cosmic rays.  In 1938, Pierre Auger and his colleagues first reported the existence of extensive air showers (EAS), showers of secondary  particles caused by the collision of primary high energy particles with air molecules. On the basis of his measurements, Auger concluded that he had observed showers with energies of $10^{15}$eV ~\cite{bib:Auger1938,bib:Auger1939}. The literature abounds in historical introductions to cosmic rays, we recommend the heart-warming notes 
by J. Cronin at the 30th International Cosmic Ray Conference ~\cite{bib:Cronin2009}. See also the lectures notes presented in \Refs ~\cite{bib:Lectures2008,bib:Lectures2011}.

\begin{figure}
\centering\includegraphics[width=.8\linewidth]{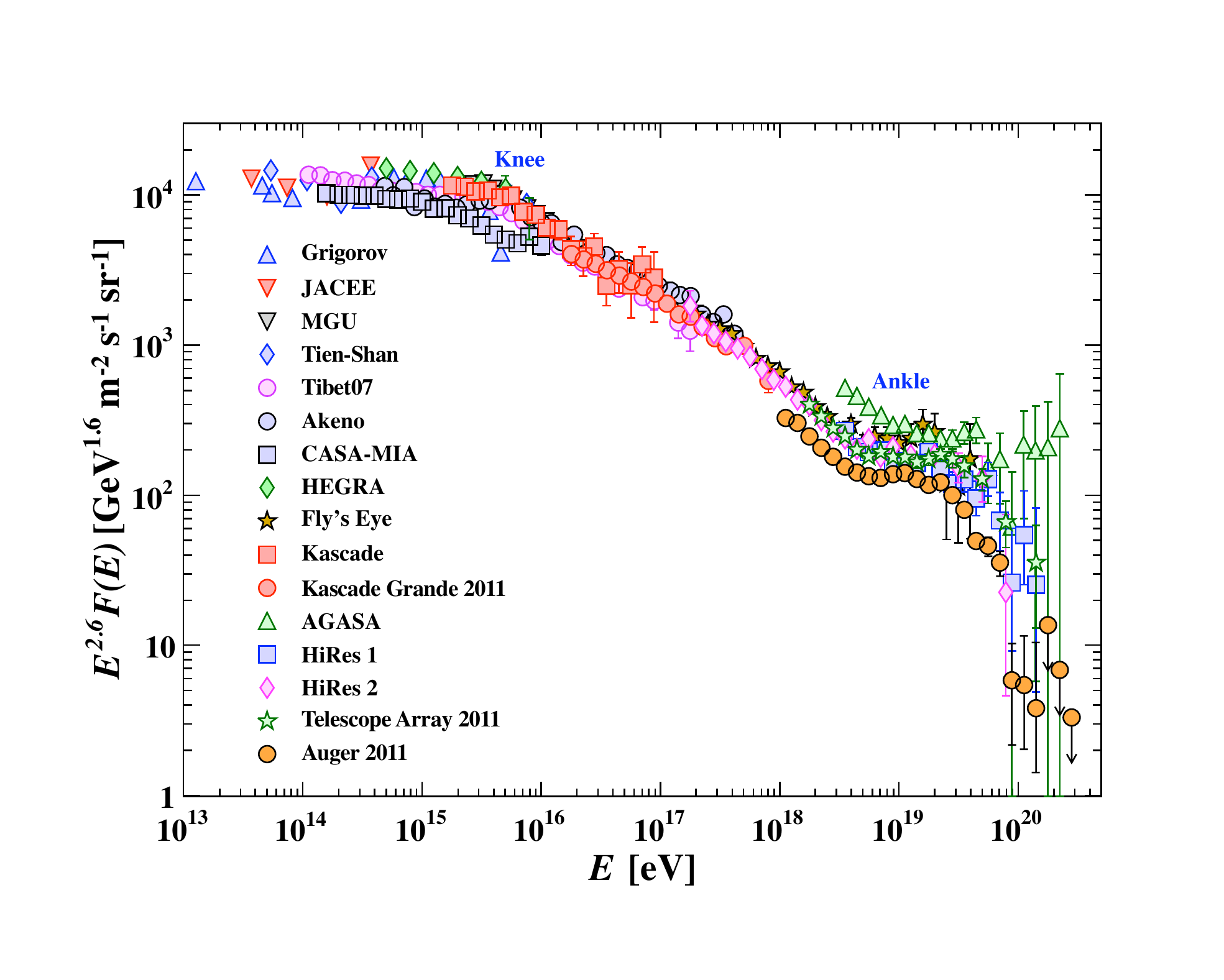}
\caption{All-particle spectrum of cosmic rays. From \Ref~\cite{bib:pdg2011}}
\label{fig:crAllParticle}
\end{figure}

For primary energy above $10^{11}$eV, the observed cosmic ray flux can be described by a series of 
power laws with the flux falling about three orders of magnitude for each decade increase in energy. Figure~\ref{fig:crAllParticle} shows the ``all-particle''  spectrum. The differential energy spectrum has
been multiplied by $E^{2.6}$ in order to display the features of the steep spectrum that are
otherwise difficult to discern~\cite{bib:pdg2011}. A change of the spectral index  (
 $E^{-2.7}$ to $E^{-3.0}$)  at an energy of about $10^{15}$eV is known as the
cosmic ray knee.  This feature is generally believed to correspond to the steepening of the galactic proton spectrum, either because a change of the propagation regime or because of maximum limitations at the source, ~\cite{bib:knee1, bib:knee2,bib:knee3}. The same effect for heavier nuclei may cause the softer spectrum above the knee. In this context, subsequent steepenings of the spectrum are predicted at $E_{max} \sim  Z \times 10^{15}$eV reaching $\sim 8  \times 10^{16}$eV for the iron group. The KASCADE-Grande collaboration provided the first observation of this sequence of changes~\cite{bib:knee4}.  Above several $\sim  10^{18}$eV the magnetic field in the vicinity of the Galaxy would not trap very effectively even the very
heaviest nuclei,  so the detected cosmic rays must be extragalactic~\cite{bib:Hillas2006}. The onset of an extragalactic contribution could be indicated by the so-called second knee, a 
further steepening of the spectrum at about $10^{17.7}$eV.   The flattening around $10^{18.5}$eV is called the ankle of
the spectrum.  
The simplest way of producing this feature is that of intersecting the steep galactic spectrum with a flatter extragalactic one. Under this assumption, several models have been developed. In the ``ankle model'' ~\cite{bib:ankle1, bib:ankle2}, the transition appears at $10^{18.5}$eV. This model needs
a new high energy galactic component between the iron knee and the onset of the extragalactic component. In the ``dip model'', the ankle appears as an intrinsic part of the  pair-production dip, a feature predicted in the spectrum of extragalactic
protons that can be directly linked to the interaction of UHECR with the CMB~\cite{bib:dip1988,bib:dip2006,bib:dip2007}.  In this model the transition from the galactic to the extragalactic component begins at the second knee and is completed at the beginning of the dip at $ E \sim 10^{18}$eV. In ``mix composition models'' ~\cite{bib:mixfew}, the  transition occurs at $3 \times 10^{18}$eV with mass composition changing from the galactic iron to extragalactic
mixed composition of different nuclei. For a recent comprehensive review of the transition models see \Ref~\cite{bib:dip2012}. 

The Large Hadron Collider (LHC) will collide in 2015 protons  at $\sqrt{s} \simeq 14$~TeV.  This impressive energy is still about a factor 
of 50 smaller than the centre-of-mass energy of the highest energy cosmic ray so 
far observed, assuming primary protons.

For cosmic ray energies above $10^{15}$eV, the flux becomes so low that direct detection
of the primary using devices in or above the upper atmosphere is, for all practical purposes, impossible. 
Fortunately, in such cases the primary 
particle has enough energy to initiate a particle cascade in the atmosphere  
large enough that the products are detectable at ground. 
There are several techniques which can be employed in detecting these extensive air showers (EAS), 
ranging from sampling of particles in the cascade
to measurements of fluorescence, \v{C}erenkov or radio emissions produced
by the shower.

\section{Acceleration of cosmic rays}

There are two types of  mechanisms able to accelerate charged particles to reach ultrahigh energies and at the same time give a power law injection spectrum. One is the acceleration of particles  directly to very high energy  by an extended electric
field~\cite{bib:ace1Hillas}, such as the case of unipolar inductors in relativistic magnetic
rotators (e.g. neutron stars~\cite{bib:ace2Blasi}) or black holes with magnetized disks that lose rotational energy in
jets. They have the advantage of being fast, however, they suffer from the circumstance that
the acceleration occurs in astrophysical sites of very high energy density, where new opportunities for
energy loss exist. In addition, they predict a hard injection spectrum that cannot be reconciled with the currently observed slope.  In 1949, Fermi  introduced a  statistical acceleration mechanism~\cite{bib:Fermi}. In his publication, Fermi considered the scattering of cosmic particles  on moving magnetized clouds which  led to a fractional energy gain  $\xi=\langle \Delta E \rangle/E \propto \beta^2$  where $\beta$ is the average velocity of the scattering centres in units of c. There is  a net transfer of the macroscopic kinetic energy from the moving cloud to the particle, but
the average energy gain is very small. Nowadays, this process is called ``second order Fermi acceleration''.  The first really successful theory of high energy cosmic ray acceleration was identified in~\cite{bib:ace3Axford} to be the Fermi acceleration in nonrelativistic shock waves in supernova remnants.  The diffusion of cosmic rays in moving magnetized plasmas in the upstream and downstream of the shocks, force particles to repeatedly cross the shock front, hence gaining energy by numerous encounters, this results in  $\xi \propto \beta$. When measured in the stationary upstream frame, $\beta$  is the speed of
the shocked fluid in units of c. This mechanism is known as ``first
order Fermi acceleration''. Shock waves  for UHECR acceleration are Gamma Ray Bursts (GRB) shocks, jets and hot spots of Active Galactic Nuclei (AGN), and gravitational accretion shocks.
  
Following~\cite{Gaisser:vg}, we provide here a simple calculation to obtain the  power law predictions from first order Fermi processes under the  ``test particle approximation'',  in which the back-reaction
of accelerated CRs on the shock properties is neglected.  The energy $E_n$ of a cosmic particle after n acceleration cycles is:
\begin{equation}
E_n = E_0  (1 + \xi)^n
\label{en}
\end{equation}
and the number of cycles to reach $E$ results from Eq.~(\ref{en})
\begin{equation}
n =   \ln \left( \frac{E}{E_0}  \right)\, / \ln (1 + \xi)
\label{n}
\end{equation}
where $E_0$ is the energy at injection into the acceleration site. If the escape probability $P_{esc}$ per encounter is constant, then the probability to stay in the
acceleration region after n cycles  is $ (1 -P_{esc})^n$. The fraction of particles accelerated to energies $>E$, the integral spectrum, is:

\begin{equation}
N(>E) \propto \frac{(1 -P_{esc})^n}{P_{esc}} \,\propto  \frac{1}{P_{esc}} \, \left( \frac{E}{E_0}  \right)^{-\gamma}
\label{spec1}
\end{equation}
with $\gamma \propto P_{esc}/ \xi$ for $ \xi \, \ll 1$ and $P_{esc}\, \ll 1$. Note that both first and second order Fermi acceleration produce a
power law energy spectrum.

The escape probability from the acceleration site depends on the characteristic time for the acceleration cycle and the characteristic time for escape from the acceleration site. In the rest frame of the shock the conservation relations imply that the upstream velocity $u_{\rm up}$ is much higher than the downstream velocity $u_{\rm down}$. The compression ratio $r= u_\text{up}/u_\text{down}=n_\text{down}/n_\text{up}$ can be determined by requiring continuity of particle number, momentum, and energy across the shock. Here  $n_\text{up}$  ($n_\text{down}$) is the particle density of the upstream (downstream) plasma.  For an ideal gas the compression ratio can be related to the specific heat ratio and the Mach number of the shock. In the case of highly supersonic shocks,  $r=4$
~\cite{bib:Blandford1987}.  To determine the spectrum we need to calculate $\gamma$. For the case of shock acceleration, $\xi=4\, \beta \,/3= 4\,(u_{\rm up} - u_{\rm down})/3 $ and the escape probability can be obtained as the ratio of the loss flux, downstream away from the shock, and the crossing flux.  Assuming the configuration of a large, plane shock the escape probability results as $P_{esc} = 4 u_{\rm down}/c$. Finally,  we obtain the spectral index of the integral energy spectrum:

\begin{equation}
\gamma \propto P_{esc}/ \xi  \propto \frac {3}{u_{\rm up} / u_{\rm down} - 1} \,\propto  1
\label{gam}
\end{equation}

This injection spectrum should be compared with the observed flux of cosmic rays,   $dN/dE \propto E^{-2}$. The result is in good agreement although additional effects, like energy losses or an energy dependence of the escape probability, could  have an important impact on the shape of  the injection spectrum.   For a comprehensive review of
shock acceleration theory, see \Ref~\cite{bib:Blandford1987}. For a discussion about different acceleration mechanisms we recommend \Ref~\cite{bib:KoteraOlinto}.

\begin{figure}
\centering
\includegraphics[width=0.5\columnwidth]{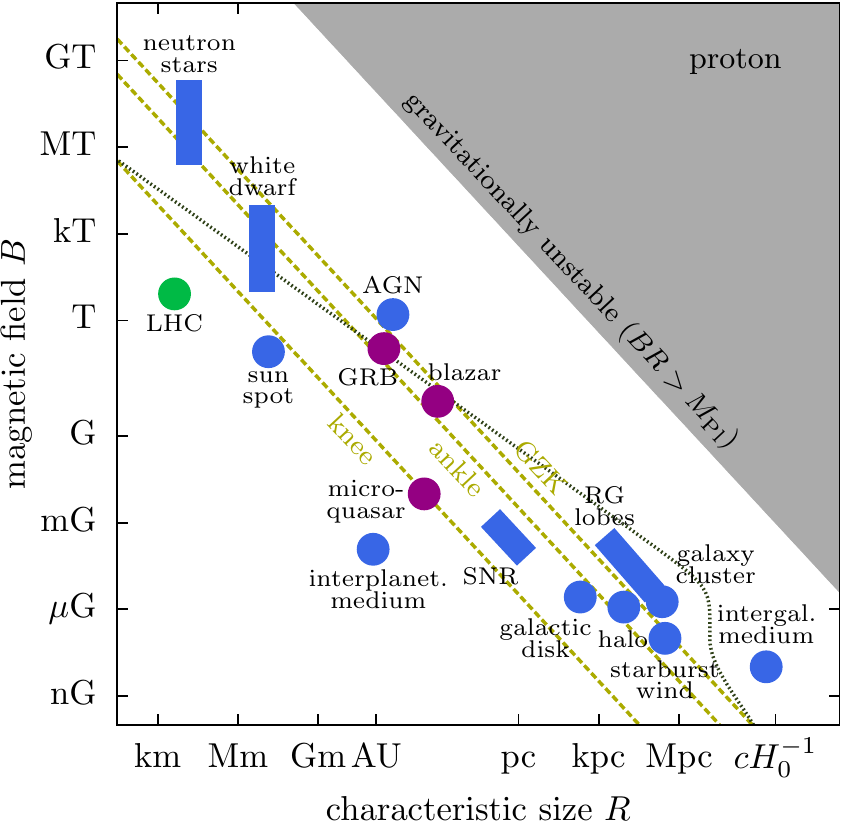}
\caption{The ``Hillas plot'' for various CR source candidates (blue). Also shown are jet-frame parameters for blazers, gamma-ray bursts, and microquasars (purple). The corresponding point for the LHC beam is also shown. The red dashed lines show the {\it lower limit} for accelerators of protons at the CR knee ($\sim 10^{14.5}$eV), CR ankle ($\sim 10^{18.5}$eV) and the GZK suppression ($\sim 10^{19.6}$eV). The dotted gray line is the {\it upper limit} from synchrotron losses and proton interactions in the cosmic photon background ($R\gg1$~Mpc). From \Ref~\cite{yellowbook}. \label{HILLASPLOT}}
\end{figure}

The requirements for astrophysical objects to be sources of UHECR are stringent. The Larmor radius of a particle with charge $Ze$ increases with its energy $E$ according to
\begin{eqnarray}\label{LARMOR}
r_L   = \frac{1.1}{Z}\left(\frac{E}{{10^{18}\text{eV}}}\right)\left(\frac{B}{\mu\text{G}}\right)^{-1}\,{\rm kpc}\,.
\end{eqnarray} 
The search for UHECR extralagalactic sources was motivated by the fact that $r_L$ in the galactic magnetic field is much larger than the thickness of the galactic disk, hence, confinement in the galaxy is not maintained for UHECR. 
The famous Hillas criteria states that the Larmor radius of the accelerated particles cannot exceed the size of the source ($R_\text{source}$), setting a natural limit in the particle's energy. 
\begin{equation}
\label{EMAX}
E_\text{max} \simeq  Z \left(\frac{B}{\mu\text{G}}\right)\left(\frac{R_\text{source}}{\text{kpc}}\right)\, \times 10^{18}~{\rm eV}\,.
\end{equation}
This limitation in energy can be seen in the so-called Hillas plot~\cite{Hillas:1985is} shown in Fig.~\ref{HILLASPLOT} where candidate sources are placed in a plane of the characteristic magnetic field $B$ versus their characteristic size $R$. For protons, the only sources for the UHECR that seem to be plausible are radio galaxy lobes and clusters of galaxies.  Exceptions may occur for sources which move relativistically in the host-galaxy frame, in particular jets from AGN and GRB. In this case the maximal energy might be increased due to a Doppler boost by a factor $\sim30$ or $\sim1000$, respectively. For a survey
of cosmic ray sources shown in Fig.~\ref{HILLASPLOT} and their signatures, see \Refs~\cite{bib:KoteraOlinto, bib:Lemoine2013}. An interesting point is that if acceleration takes place in GRB, one may expect a strong neutrino signature due to proton interactions with the radiative
background~\cite{bib:Waxman1997}. Such a signature is now being probed by the Ice Cube experiment~\cite{bib:IceCubeNature}.

\section{Propagation of extragalactic cosmic rays}
\subsection{Energy losses of protons}

There are three main energy loss processes for protons propagating over cosmological distances: Adiabatic energy losses due to the expansion of the universe,  $-dE/dt = H_0 $, 
pair 
production $(p \gamma \rightarrow p  e^+  e^-$)  and pion-production  $p \gamma \rightarrow \pi  N$ on photons of the cosmic microwave background (CMB).  Collisions with optical and infrared photons give a negligible contribution.

The fractional  energy loss due to interactions with the cosmic background radiation at a redshift $z=0$ is determined by the integral of the nucleon  energy loss per collision multiplied by the probability per unit time for a nucleon collision in an isotropic gas of photons~\cite{Stecker:68}.  For interactions with a blackbody field of temperature $T$, the photon 
density is that of a Planck spectrum, so the 
fractional energy loss is given by
\begin{equation}
-\frac{1}{E} \frac{dE}{dt} = - \frac{ckT}{2 \pi^2 \Gamma^2 (c \hbar)^3}
\sum_j \int_{\omega_{0_j}}^{\infty}  d\omega_r \,
\sigma_j (\omega_r) \,y_j \, \omega_r \, \ln ( 1 -  e^{-\omega_r / 2 \Gamma kT}) \,,
\label{phds!}
\end{equation}
where $\omega_r$ is the photon energy in the rest frame of the nucleon, and $y_j$ is the inelasticity, {\em i.e.}  the average fraction of the energy lost by the photon to the nucleon in the laboratory frame for the $j$th reaction channel. The sum is carried out over all channels and $d\omega$, $\sigma_j(\omega_r)$ is the total cross section of the $j$th interaction channel, $\Gamma$ is the usual Lorentz factor of the nucleon, and  $\omega_{0_j}$ is the threshold energy for the $j$th reaction 
in the rest frame of the nucleon.

At energies $E\ll m_e\,m_p/kT = 2.1 \times 10^{18}$eV, the reaction $(p \gamma \rightarrow p  e^+  e^-$) takes place on the photons 
from the high energy tail of the 
Planck distribution. The cross section of the reaction approximated by the threshold values is $\sigma(\omega_r) = \frac{\pi}{12}\, \alpha\,\, r_0^2  \left(\frac{\omega_r}{m_e} - 2\right)^3\,$,  $\alpha$ is the fine structure constant  and $r_0$ is the classical radius of the electron~\cite{Berezinsky:wi}. The inelasticity at threshold results $y_{_{\rm }} = 2\,\frac{m_e}{m_p}\,$. 
The fractional energy loss due to pair production is 
then, 
\begin{equation}
-\frac{1}{E}\, \left(\frac{dE}{dt}\right)_{\rm } = 
\frac{16 c}{\pi}\, \frac{m_e}{m_p}\, 
\alpha\, r_0^2\,
\left(\frac{kT}{hc}\right)^3 \, \left(\frac{\Gamma k T}{m_e}\right)^2\, 
\exp \left(-\frac{m_e}{\Gamma kT}\right).
\label{uniden}
\end{equation}
At higher energies ($E>10^{19}$eV) the photopion 
reactions $p \gamma \rightarrow p \pi^0$ and
$p \gamma \rightarrow \pi^+  n$ on the tail of the Planck distribution give 
the main contribution to proton energy loss. The photons are seen blue-shifted by the cosmic rays in their rest frames and the reaction becomes possible. The cross sections of these 
reactions are well known. It strongly increase at the $\Delta (1232)$ 
resonance, which decays into the one pion channels $\pi^+ n$ and $\pi^0 p$ at a photon energy in the proton rest frame of 
145~MeV. At higher energies, heavier baryon resonances occur and the proton 
might reappear only after successive decays of resonances. The cross section in this 
region can be described by a sum of Breit-Wigner 
distributions over the main 
resonances produced in $N \gamma$ collisions with $\pi N$, 
$\pi \pi N$ and $K\Lambda$ ($\Lambda \rightarrow N \pi$) final 
states~\cite{Barnett:1996hr}.  For the cross section at high energies the fits from the CERN-HERA and COMPAS Groups to
the high-energy  $p\gamma$ cross section~\cite{Montanet:1994xu} can be used.  Assuming that reactions mediated by baryon resonances have spherically symmetric decay angular distributions, the average energy loss of the nucleon after $n$ resonant collisions is given by 
\begin{equation}
y_\pi(m_{R_0}) = 1 - \frac{1}{2^n} \prod_{i=1}^{n} \left( 1 + \frac{m_{R_{_i}}^2 -
m_M^2}{m_{R_{_{i-1}}}^2} \right)\,,
\label{kj}
\end{equation}
where $m_{R_{_i}}$ denotes the mass of the $i^{\rm th}$ resonant system of the decay chain, $m_M$ the mass of the associated meson, $m_{R_{_0}} = \sqrt{s}$ is the total energy of the reaction in the c.m., and $m_{R_{_n}}$ the mass of the nucleon.  It is well established from experiments that, at very
high energies ($\sqrt{s}>3$~GeV), the incident nucleons lose one-half their energy via pion
photoproduction independent of the number of pions produced ( ``leading particle effect'')~\cite{Golyak:cz}.

\begin{figure}
\centering
\includegraphics[width=0.5\columnwidth]{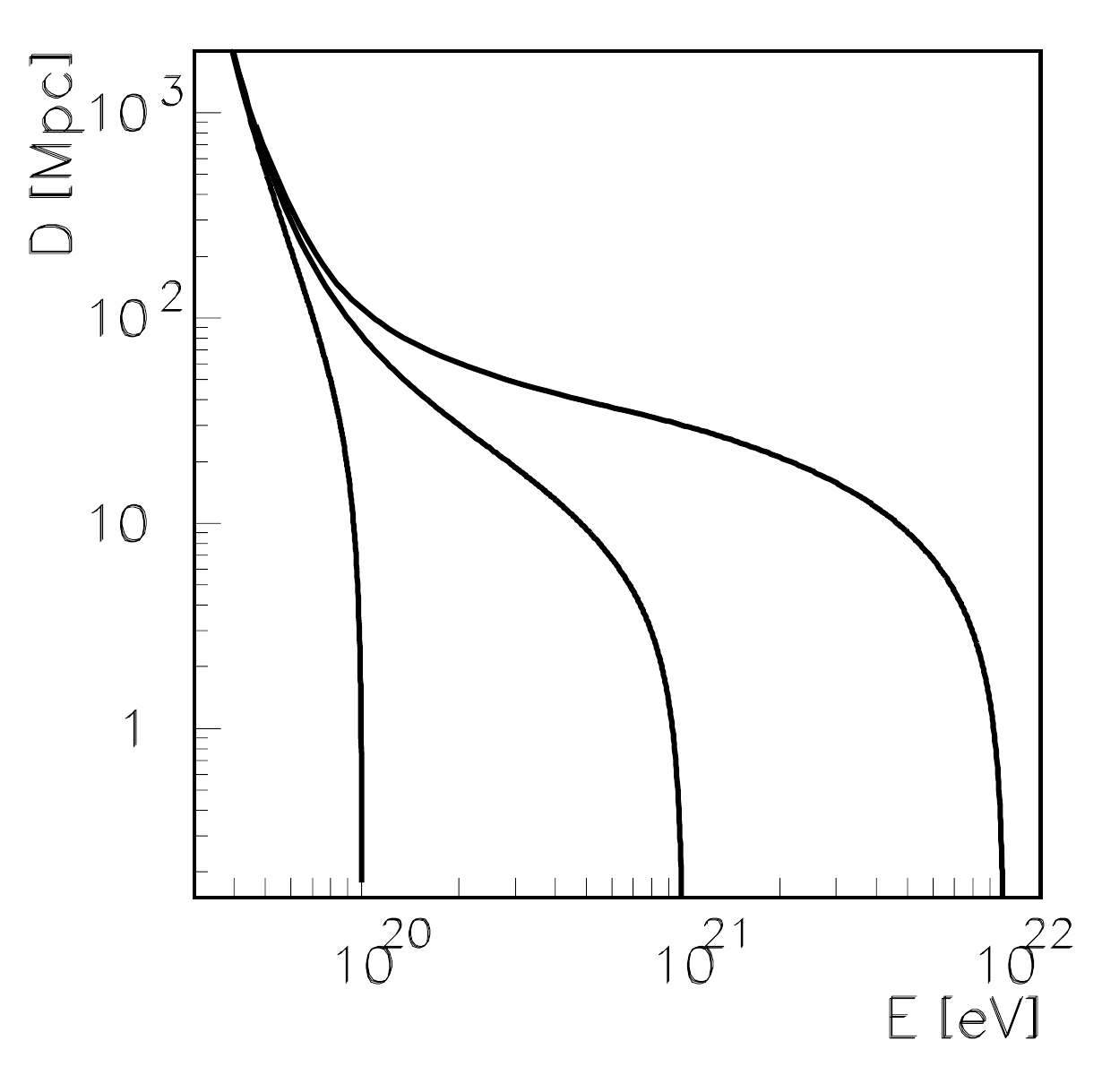}
\caption{Energy attenuation length of protons in the intergalactic
medium. For  proton sources beyond $\approx 100~{\rm Mpc}$, the observed proton energy is $<10^{20}$eV regardless its initial value. 
From \Ref~\cite{Anchordoqui:1996ru}.} 
\label{gzk}
\end{figure}

A fit to Eq.~(\ref{phds!}) for the region $\sqrt{s} < 2$ GeV with the exponential behavior derived from the values of cross section 
and fractional energy loss at threshold, gives~\cite{Anchordoqui:1996ru} 
\begin{equation} 
- \frac{1}{E}\,\left(\frac{dE}{dt}\right)_\pi = A \, {\rm exp} [ - B / E ]\,,
\label{okop}
\end{equation}

\begin{equation}
A = ( 3.66  \pm 0.08 )
\times 10^{-8} \, {\rm yr}^{-1}, \,\,\,\,\, B = (2.87 \pm 0.03 )\times
10^{11}\, {\rm GeV} \, .
\label{parame}
\end{equation} 
The fractional energy loss at higher c.m. energies ($\sqrt{s} \gtrsim 3$~GeV) 
is roughly a constant, 
\begin{equation}
-\frac{1}{E}\,\left(\frac{dE}{dt}\right)_\pi = C = ( 2.42 \pm 0.03 ) 
\times 10^{-8}  \,\, {\rm yr}^{-1} \,.
\label{ce}
\end{equation}
From the values determined for the fractional energy loss, it is 
straightforward to compute the energy degradation of UHECRs in terms of their flight time. This is given by,
\begin{equation}
A \, t \, - \,  {\rm Ei}\,(B/E) 
+ \, {\rm Ei}\, (B/E_0) 
= 0\,,\,\,\,\,\,\,  {\rm for} \,\,10^{10}\,{\rm GeV} \lesssim E \lesssim 10^{12} \,{\rm GeV} \,,
\label{degradacion}
\end{equation}
and
\begin{equation}
E (t) = E_0 \exp[- \,C \ t \,]\,,\,\,\,\,\,\, {\rm for}\,\, E \gtrsim 10^{12} \, 
{\rm GeV}\,,
\end{equation}
where Ei is the exponential integral. Figure~\ref{gzk} shows the proton energy degradation as a function of the mean propagation distance. Notice that, independent of the initial energy of the nucleon, the mean energy values approach $10^{20}$eV after a distance of $\approx 100~{\rm Mpc}$. This fact contrains the proximity to the Earth of the sources of UHECR with energies above $5 \,\times10^{19}$eV. 

\subsection{Energy losses of nuclei}

The relevant mechanisms for the energy loss of nuclei during propagation are: Compton interactions, pair production in the field of the nucleus, photodisintegration and hadron photoproduction. For nuclei of energy 
$E>10^{19}$eV the  dominant loss process is photodisintegration.  In the nucleus rest-frame, pair production has a threshold at $\sim 1$~MeV, photodisintegration is particularly important at the peak of the giant dipole resonance (15 to 25~MeV), and photomeson production has a threshold energy of $\sim 145$~MeV. Compton interactions result in only a negligibly small energy loss for 
the nucleus~\cite{Puget:nz}.

For a nucleus of mass $A$ and charge $Ze$, the energy loss rate due to 
photopair production is $Z^2/A$ 
times higher than for a proton of the same Lorentz factor~\cite{Chodorowski}, 
whereas the 
energy loss rate due to photomeson production remains roughly the same. 
The latter is true because the cross section for photomeson production by 
nuclei is proportional to the mass number $A$~\cite{Michalowski:eg}, 
while the inelasticity is 
proportional to $1/A$. However, it is photodisintegration rather than 
photopair and photomeson production that determines the energetics of 
ultrahigh energy cosmic nuclei. During this process some fragments of 
the nuclei are released, mostly single neutrons and protons.  
Experimental data of photonuclear interactions are consistent with a 
two-step process: photoabsorption by the nucleus to form a compound state, 
followed by a statistical decay process involving the emission of one or more 
nucleons.

The disintegration rate with production of $i$ nucleons is given 
by~\cite{Stecker:fw}
\begin{equation}
R_{Ai} = \frac{1}{2 \Gamma^2} \int_0^{\infty} dw \,
\frac{n(w)}{w^2} \, \int_0^{2\Gamma w} dw_r
 \, w_r \sigma_{Ai}(w_r)
\label{phdsrate}
\end{equation}
where $n(w)$ is the density of photons with energy $w$ in the 
system of reference in which the cosmic microwave background (CMB) 
is at 2.7~K and
$w_r$ is the energy of the photons in the rest frame of the nucleus.
As usual, $\Gamma$ is the Lorentz factor and $\sigma_{Ai}$
is the cross section for the interaction. 

Here, the soft photon background is taken as the sum of a 2.7~K Planckian 
spectrum that dominates at energies 
$w \in (2.0 \times 10^{-6}~{\rm eV}\,, 4 \times 10^{-3}~{\rm eV})$, and 
the infrared 
radiation as estimated in \Ref~\cite{Malkan:1997yd}.
Parameterizations 
of the photodisintegration cross section for the different
nuclear species are given in \Ref~\cite{Puget:nz}. Summing over all 
possible channels for a given 
number of nucleons, one obtains the effective nucleon loss rate 
$R = \sum_i i R_{Ai}.$ The effective nucleon loss rate for light 
elements, as well as for those in  the carbon, silicon 
and iron groups can be scaled as in~\cite{Puget:nz}
\begin{equation}
\left. \frac{dA}{dt}\right|_A \sim \left. 
\frac{dA}{dt}\right|_{\rm Fe} \left(\frac{A}{56}\right) = 
\left. R\right|_{_{\rm Fe}} \left(\frac{A}{56}\right)\,,
\end{equation}
with the photodisintegration rate parametrized by~\cite{Anchordoqui:1997rn}
\begin{equation}
R _{56} (\Gamma) =3.25 \times 10^{-6}\, 
\Gamma^{-0.643}                                          
\exp (-2.15 \times 10^{10}/\Gamma)\,\, {\rm s}^{-1} 
\label{oop}
\end{equation}
for $\Gamma \,\in \, [1.0 \times 10^{9}, 36.8 \times 10^{9}]$, and 
\begin{equation}
R_{56}(\Gamma) =1.59 \times 10^{-12} \, 
\Gamma^{-0.0698}\,\, {\rm s}^{-1}   
\end{equation}
for $ \Gamma\, 
\in\, 
[3.68 \times 10^{10}, 10.0 \times 10^{10}]$. \\

For photodisintegration, the averaged fractional
energy loss results equal to the fractional loss in mass number of the 
nucleus, because the nucleon emission is isotropic in the
rest frame of the nucleus. During the photodisintegration process the 
Lorentz factor of the nucleus is conserved, unlike the cases of pair 
production and photomeson production processes which involve the creation 
of new 
particles that carry off energy. The total fractional energy loss 
 is then 
\begin{equation}
-\frac{1}{E} \frac{dE}{dt} = \frac{1}{\Gamma} \frac{d\Gamma}{dt} + \frac{R}{A} \,.
\end{equation}
For $\omega_r \lesssim 145$~MeV the reduction in $\Gamma$ comes from the nuclear 
energy loss due to pair production~\cite{Stecker:1998ib}.
For $\Gamma > 10^{10}$ the  energy loss due to photopair production is 
negligible, and thus
\begin{equation}
E (t)  \sim  938\,\, A(t)\,\, \Gamma\,\,\, {\rm MeV}  \,\,
   \sim  E_0\, e^{- \left. R(\Gamma)\right|_{_{\rm Fe}}\,t/56}\,.
\label{pats}
\end{equation}

\begin{figure}
\begin{center}
\includegraphics[height=6.cm]{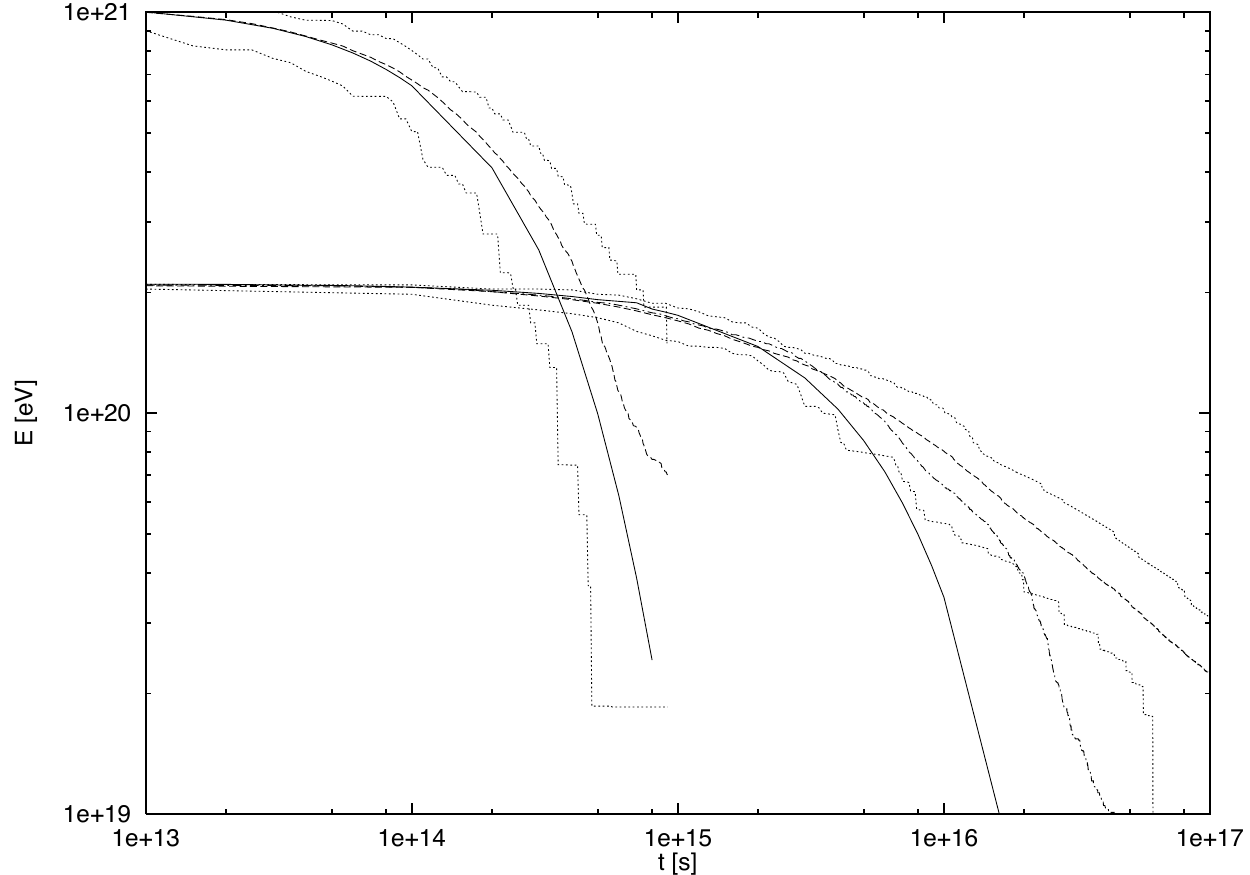}
\caption{The energy of the surviving fragment ($\Gamma_0 = 4 \times
10^{9}$, $\Gamma_0 = 2 \times 10^{10}$) vs. propagation time
obtained using
Eq.~(\ref{pats}) is indicated with a solid line.
Also included is the energy attenuation
length obtained from Monte Carlo simulations with (dashed) 
and without (dotted-dashed) pair creation production, for comparison.
The region between the two dotted lines includes 95\% of the 
simulations. This
gives a clear idea of the range of values which can result from
fluctuations from the average behaviour.} 
\label{gzk2}
\end{center}
\end{figure}

Figure~\ref{gzk2} shows the energy of the heaviest surviving nuclear
fragment as a function of the propagation time, for initial iron nuclei. 
The solid curves are obtained using Eq.~(\ref{pats}), whereas
the dashed and dotted-dashed curves are obtained by means of
Monte Carlo simulations~\cite{Epele:1998ia}. One can see that nuclei with 
Lorentz factors above $10^{10}$ cannot survive 
for more than 10~Mpc. For these distances, the approximation given in 
Eq.~(\ref{pats}) always lies in the region which includes 95\% of the 
Monte Carlo simulations. When the nucleus is emitted with a Lorentz factor 
$\Gamma_0 < 5 \times 10^9$, pair production
losses start to be relevant, significantly reducing the value of $\Gamma$ 
as the nucleus propagates distances of ${\cal O}$(100~Mpc). 
The effect has a maximum for $\Gamma_0 \approx 4 \times 10^{9}$
but becomes small again for $\Gamma_0 \leq 10^{9}$, for which
appreciable effects only appear for cosmological distances ($>1000$
Mpc), see for instance \Ref~\cite{Epele:1998ia}. 

Note that Eq.~(\ref{pats}) imposes a  strong 
constraint on the location of nucleus-sources:  less than 1\%
of iron nuclei (or any surviving fragment of their spallations)
can survive more than $3 \times 10^{14}$~s with an energy $>10^{20.5}$ eV. It is important to keep in mind that
a light propagation distance of $1.03 \times 10^{14}$~s corresponds to 
1~Mpc.

In recent years the interest in the propagation
of UHECR nuclei has significantly grown. A complete review with a detailed list of references can be found in ~\cite{Allard:2011aa}. Most recent calculations of  UHECR proton propagation use the Monte Carlo generator SOPHIA~\cite{SOPHIA} for photomeson interaction of protons, based on available data and 
phenomenological models. For the case of nuclei propagation, existing propagation codes are CRPropa~\cite{CRPropa} and the complete nuclei propagation tool presented in \Ref~\cite{Allard:2006}.

\section{Cosmic ray observations at the highest energies: Hybrid instruments}

For primary cosmic ray energies above $10^{14}$eV, the flux becomes so low that individual events cannot longer be detected directly.  
Fortunately, in such cases the primary 
particle has enough energy to initiate an extended air shower (EAS)  in the atmosphere. Only the secondary particles are detected and used to infer the properties of the primary particle.  There are several techniques which can be employed in detecting EAS.

The most commonly used detection method involves sampling the shower 
front at a given altitude using an array of sensors spread over a 
large area.  The classical set up consists of an array of plastic
scintillators, registering charged particles from the shower (also some converted photons). Another technique is to use water  \v{C}erenkov detectors (WCD), that  allow the detection of the very numerous photons present in showers. They are deep compare with scintillators, so they have larger response to inclined showers.   An initial estimate of the shower direction is obtained from the relative
arrival times of signal at a minimum of 3 non-collinear detectors, treating
the shower front as if it were planar. The density of particles falls off with the distance to the shower core and this can be parameterized by a lateral distribution function (LDF), which, of course, depends on the characteristics of the detectors used.  The particle density at a large distance 
from the shower core is commonly used as an energy estimator. Muons in the EAS have higher 
energies than electromagnetic particles, which in addition suffer 
significant scattering and energy loss. Thus, the muonic component tends 
to arrive earlier and 
over a shorter period of time than the electromagnetic one. These
signatures may also  help to distinguish $\mu $'s from electrons and $\gamma$'s providing a 
useful tool to determine the primary composition.

Another highly successful air shower detection method involves measurement of the longitudinal
development of the cascade by sensing the fluorescence light produced
via interactions of the charged particles in the atmosphere.  
As an extensive air shower develops, it dissipates much of its energy by
exciting and ionizing air molecules along its path.  Excited nitrogen
molecules fluoresce producing radiation in the 300 - 400~nm ultraviolet
range, to which the atmosphere is quite transparent. Under favourable
atmospheric conditions EAS can be detected at distances as large as
20~km,  though observations can only be made on clear moonless nights, yielding a 
duty cycle of about 10\%.  The shower development appears as a rapidly moving spot of light
whose angular motion depends on both the distance and the orientation of the shower axis.
  The fluorescence technique provides the most effective way to
measure the energy
of the primary particle. The amount of fluorescence light emitted is
proportional to
the number of charged particles in the showers allowing a direct
measurement of
the longitudinal development of the EAS in the atmosphere. For this, the sky is viewed by many segmented eyes using photomultipliers.
From the measured shower profile the position of the shower maximum $X_{max}$, which is sensitive to primary composition,  can be
obtained. The energy in the electromagnetic component is calculated by
integrating the measured shower profile, after corrections for atmospheric attenuation of the fluorescence light and contamination of the signal by \u{C}erenkov light. Finally, to derived the total energy of the shower, an estimate of the missing energy carried to the ground by neutrinos and high energy muons must be made based on assumptions about the primary mass and the appropriate hadronic interaction models.

In this note we focus on the two high energy cosmic ray experiments currently operating: the Pierre Auger Observatory~\cite{Abraham:2004dt} and the Telescope Array (TA)~\cite{TA2012}. The Pierre Auger Observatory, the largest UHECR experiment in the world, is located in Malarg\"ue, Argentina ($35^\circ 12'$S, $69^\circ 12'$W).  It has an accumulated  exposure of about
$30 000 \, \rm{km}^2$ sr yr. The Telescope Array  located in Millard County, Utah, USA ($39.3^\circ N, 112.9^\circ$W),  due to a later start and its more than 4 times
smaller area, has collected about 10 times less events.  Both the Pierre Auger Observatory and TA are hybrid detectors employing
two complementary detection techniques for the
ground-based measurement of air showers induced by UHECR:
a surface detector array (SD) and a fluorescence detector
(FD).

The ground array of the Pierre Auger Observatory consists of 1600 stations
spaced by 1.5 km covering an area of 3000 km$^2$.. Each detector is a cylindrical, opaque  tank of 10 m$^2$ and a water depth of 1.2 m, where particles 
produce light by \u{C}erenkov radiation. The filtered water is contained in an 
internal coating which diffusely reflects the light collected by three
photomultipliers (PMT) installed on the top. The large diameter PMTs ($\approx$ 20~cm ) hemispherical photomultiplier are mounted facing down and look at the water through sealed polyethylene  windows that are integral part of the internal liner.   Due to the size of the array the stations have to work in an autonomous way. Thus the stations operate on battery-backed solar power 
and  communicate with a central station by using wireless LAN radio links. The time information is obtained from the Global Positioning Satellite (GPS) system.  This array is fully efficient at energies above $E>3 \times10^{18}$eV. Additional detectors with 750~m spacing have been nested
within the 1500~m array to cover an area of $25\, \rm{km}^2$ with
full efficiency above $E>3 \times10^{17}$eV. The SD is sensitive
to electromagnetic and muonic secondary particles of
air showers and has a duty cycle of almost 100\%.
The surface array is overlooked by 27 optical telescopes grouped in
5 buildings on the periphery of the array~\cite{Abraham:2010}. The field of view
of each telescope is $30\,^{\circ}$ in azimuth, and $1.5\,^{\circ}$ to $30\,^{\circ}$ in
elevation, except for three of them, for which the elevation
is between  $30\,^{\circ}$ and $60\,^{\circ}$ (HEAT telescopes~\cite{HEAT} ). Light is
focused with a spherical mirror of $13\, \rm{m}^2$ on a camera of
440 hexagonal PMTs. The FD can only operate
during dark nights, which limits its duty cycle to 13\%.  Stable data taking with
the SD started in January 2004 and the Observatory has
been running with its full configuration since 2008. 

\begin{figure}[t]
\begin{center}
\includegraphics[width=0.5\textwidth]{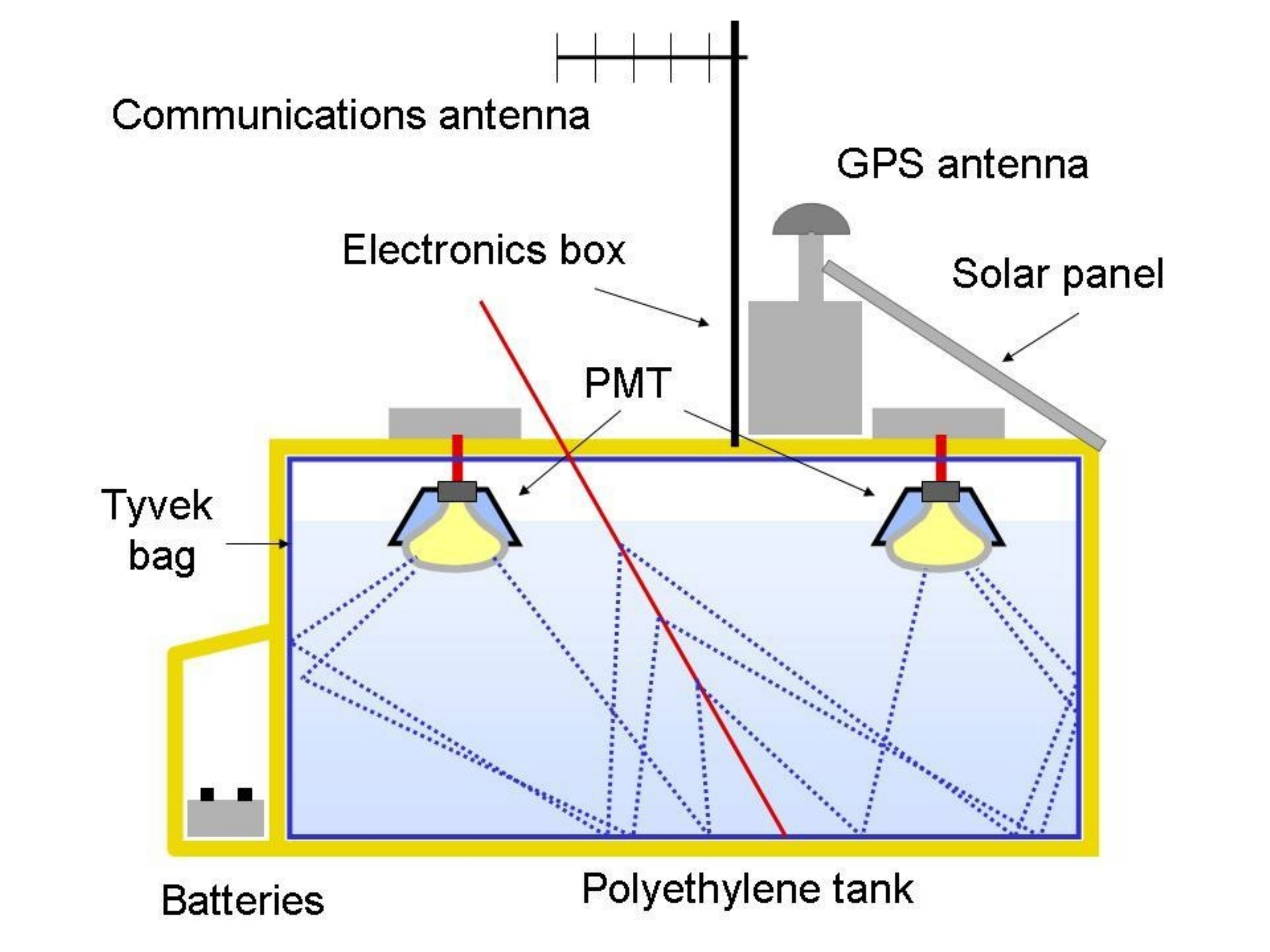}
\includegraphics[width=0.38\textwidth]{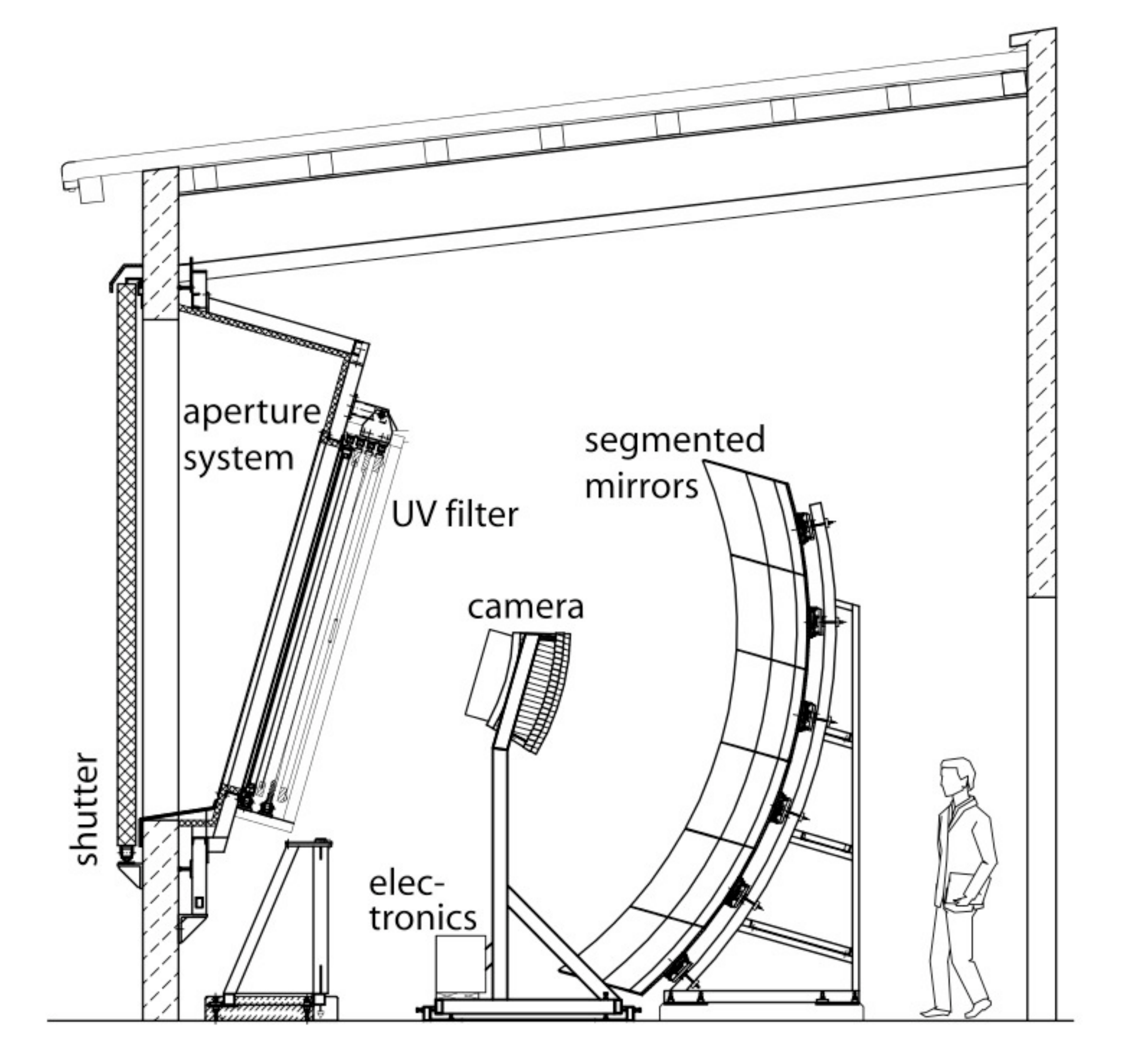}
\end{center}
\caption{Left: A typical surface detector of the Auger Observatory.   Right: A  fluorescence telescope. See the text for the description of the components.}
\label{detectors}
\end{figure}

In Figure~\ref{detectors} (left panel) we present a schematic description of a water \u{C}erenkov detector installed  at the Pierre Auger Observatory. Mounted on top of the tank are 
the solar panel, electronic enclosure, mast, radio antenna and GPS 
antenna for absolute and relative timing. A battery is contained in a box attached to the the tank.  The main components of a fluorescence eye are shown on the right panel of Figure ~\ref{detectors}: a large spherical mirror with a radius of curvature of 3.4 m, a pixel camera  in the focal surface and a diaphragm with an entrance glass
window. This filter allows reduction of night background with respect to the fluorescence signal and also serves to protect the equipment from dust. 

 The TA surface array consists of 507 detector units  deployed in a square grid with 1.2~km spacing to cover a total area of approximately 700 km$^2$. Each unit consists of a plastic scintillation counter of $3 \, \rm{m}^2$ surface and 1.2~cm thickness, with 2 layers of plastic scintillators viewed by PMT at each end. The entire system is
powered by a solar panel and battery. The communication is done with WLAN modem.  The SD array is fully efficient for cosmic rays
with energies greater than  $10^{18.8}$eV~\cite{TAeff}. Three FD stations are placed around the SD array, with a total of 38 telescopes.  Each telescope is comprised of a cluster of photo-tubes and a reflecting mirror of 3.3~m diameter.  A PMT camera consisting of $16\times16$ PMTs is set at a distance of 3000~mm from the mirror. The field of view
of each PMT is approximately $1\,^{\circ}$ and that of the FD station is from $3\,^{\circ}$ to $33\,^{\circ}$ in elevation and $108\,^{\circ}$ in azimuth.  See \Ref~\cite{TA2012}
 for details of the TA
detectors.

\section{Flux measurements}

Surface arrays, with its near 100\% duty cycle, give the larger data sample used to obtain the energy spectrum. 
The comparison of the shower energy, measured using fluorescence, with the SD energy parameter  for a subset of hybrid events is used to
calibrate the energy scale for the array. 

\begin{figure}[t]
\begin{center}
\includegraphics[width=0.39\textwidth]{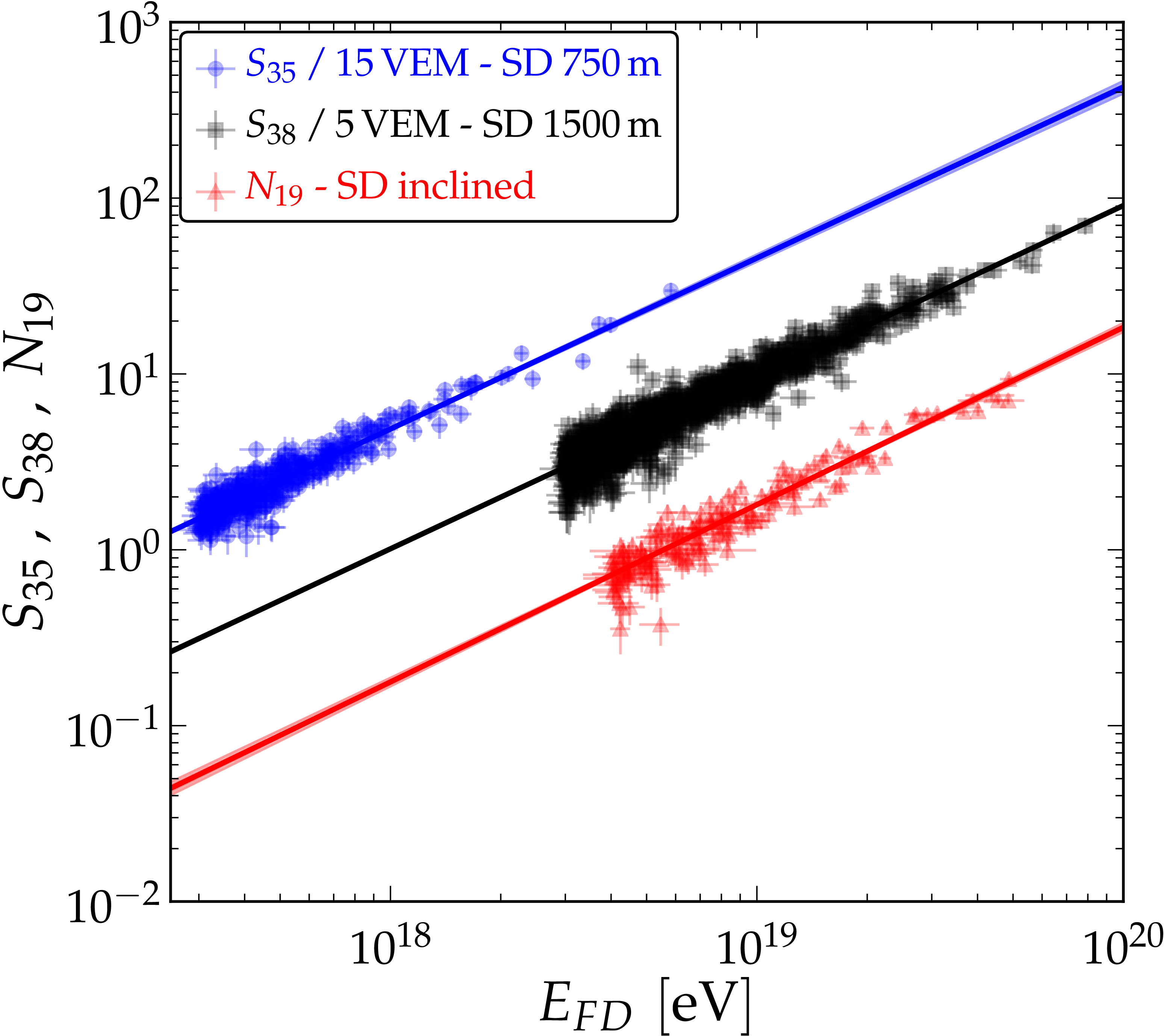}
\includegraphics[width=0.51\textwidth]{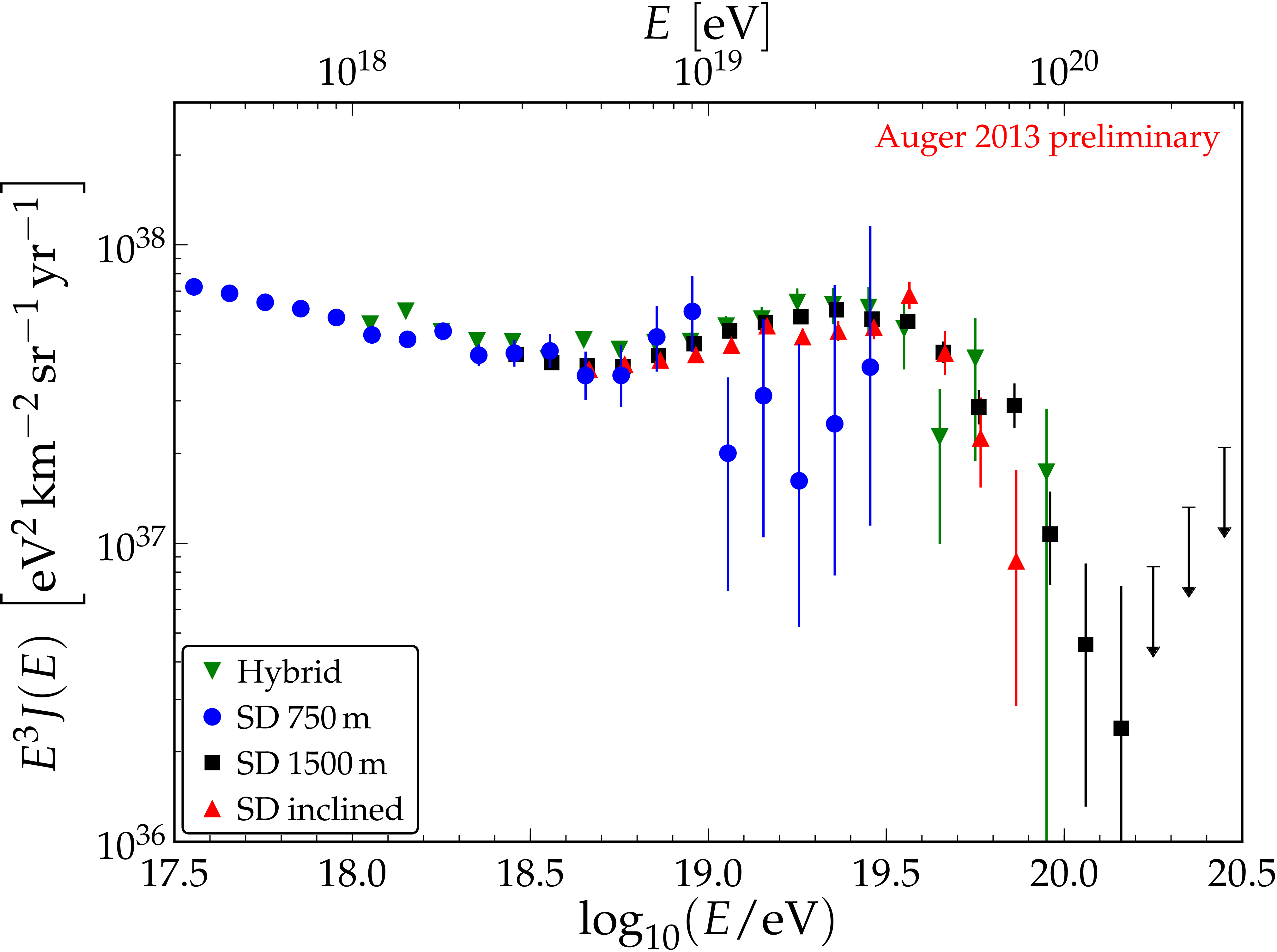}
\end{center}
\caption{Left: The correlation between the different energy estimators S38, S35 and N19 (see text) and the energy determined by FD. Right: Energy spectra, corrected for energy resolution, derived from SD and from hybrid data. From \Ref~\cite{Auger:ICRC1}. }
\label{correlation}
\end{figure}

The first step towards the flux measurement with the SD array is the reconstruction of arrival direction and core position
of air showers. Then, a stable parameter from the SD which correlates with the primary energy is reconstructed. This parameter is the signal at an optimal distances to the shower core 
at which the spread in the signal size is minimum~\cite{HillasSD}. In the following we distinguish between \textit{vertical events} ($\theta < 60^\circ$) and \textit{inclined events} ($62^\circ \leq \theta < 80^\circ$).
For the case of Auger, the optimal distance is 1000~m for the main array and 
450~m for the ``infill'', while for TA is 800~m. For \textit{vertical events} the signals at the optimal distance obtained from a LDF  fit, have to be corrected for their zenith angle
dependence due to air shower attenuation in the atmosphere. This is done in Auger with a Constant Intensity Cut (CIC) method~\cite{CIC}. The
equivalent signal at median zenith angle of  $38\,^{\circ}$ ($35\,^{\circ}$) is then used
to infer the energy for the 1500~m (750~m) array~\cite{Auger:ICRC1,CIC2}. Events that have independently triggered the SD array
and FD telescopes are used
for the energy calibration of SD data~\cite{Pesce}. The correlation between the different energy estimators
and the energy obtained from the FD is shown in Figure~\ref{correlation} (left panel) superimposed with the calibration
functions resulting from maximum-likelihood fits. For the case of TA, the energy is estimated by using a look-up
table in S(800) and zenith angle determined from an exhaustive Monte Carlo simulation. The uncertainty in energy scale of the Monte Carlo simulation of an SD is large,
and possible biases associated with the modelling of hadronic interactions  are difficult to determine. Therefore, the SD energy scale is corrected to the TA FD using hybrid events. The observed
differences between the FD and SD events are well described by a simple proportionality
relationship, where the SD energy scale is 27\% higher than the FD~\cite{TAenergy}.

Water  \v{C}erenkov detectors from the Pierre Auger Observatory SD,  have larger response to inclined showers. These EAS are characterized by the dominance
of secondary muons at ground, as the electromagnetic component
is largely absorbed in the large atmospheric depth
traversed by the shower~\cite{Auger:inclined}. The reconstruction is based on
the estimation of the relative muon content N19 with respect
to a simulated proton shower with energy $10 \times10^{19}$eV~\cite{Auger:inclined2}. N19
is used to infer the primary energy for inclined events, as shown in the left pannel of Figure~\ref{correlation}.

The energy spectra obtained from the three SD datasets
are shown in the right panel of Figure~\ref{correlation}. To characterize the spectral features, the Auger collaboration describes the data with a power law below the ankle $J(E) \propto E^{-\gamma_1}$ and a power law with smooth suppression above:
\begin{eqnarray*}
J(E; E > E_a) \propto E^{-\gamma_2} \left[ 1 + \exp{ \left( \frac{\log_{10} E - \log_{10} E_{1/2}}{\log_{10} W_c} \right) } \right]^{-1}.
\end{eqnarray*}
$\gamma_1$, $\gamma_2$ are the spectral indices below/above the ankle at $E_a$. $E_{1/2}$ is the energy at which the flux has dropped to half of its peak value before the suppression, the steepness of which is described with $\log_{10} W_c$. The data in Figure~\ref{correlation} clearly exhibit the ankle at $10^{18.7}$eV   and a flux suppression above $10^{19.6}$eV. The Pierre Auger Observatory has confirmed the GZK feature of the spectrum with a significance greater than 20 $\sigma$ obtained by comparison to a power
law extrapolation. This observation seems to indicate that acceleration in extragalactic sources can explain the high energy CR spectrum, ending the need for exotic alternatives designed to avoid the flux suppression. However, the possibility that this feature in the spectrum is due to the maximum energy of acceleration at the sources is not easily dismissed. 

We present here only the energy spectrum from the Pierre Auger Observatory, details of the corresponding spectrum obtained by the Telescope Array collaboration are presented in \Ref~\cite{TA:spectrum}.  As discussed in \Ref~\cite{TAvsAuger}, it is found that the energy spectra determined by these experiments are consistent in normalization and shape after energy scaling factors  are
applied. Those scaling factors are within systematic uncertainties in the energy scale
quoted by the experiments. 

\section{Correlation with astrophysical objects}

Since the UHECR are charged particles, they not only lose energy in the interaction with background photons, but also they are deflected by galactic and extragalactic magnetic fields. The galactic magnetic field (GMF) can be modelled as the sum of a regular (large scale fluctuations) and a turbulent (smaller scale fluctuations) components. The directions on the sky in which cosmic rays are deflected strongly depend on the GMF model, however, averaged quantities
such as the average UHECR deflection angle are much less model dependent~\cite{Giacinti:2013yya}. Extragalactic magnetic fields are expected to be stronger in the large scale structure of the Universe
and significantly weaker in voids. UHECR deflections in such fields are poorly constrained ranging from negligible to more than ten degrees, even for 100 EeV protons (See \Ref~\cite{bib:KoteraOlinto} and references therein). Attempts to detect anisotropies at ultrahigh energies are based on the selection of events with the largest magnetic rigidity to study whether they can be correlated with the direction of possible sources or distributions of astrophysical objects in our vicinity (less than 100~Mpc).  

\begin{figure}
\centering
\includegraphics[width=.5\linewidth]{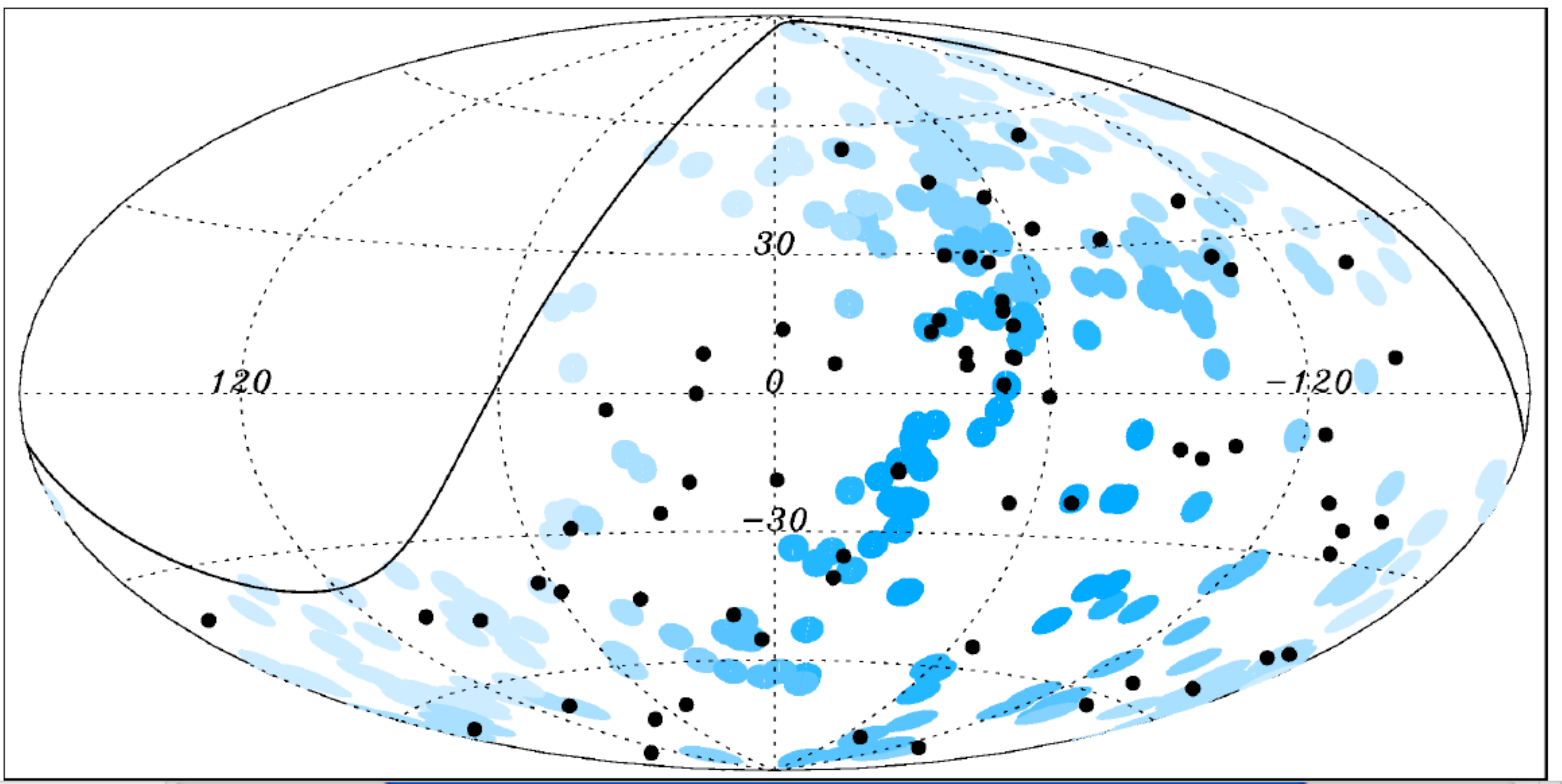}
\includegraphics[width=.44\linewidth]{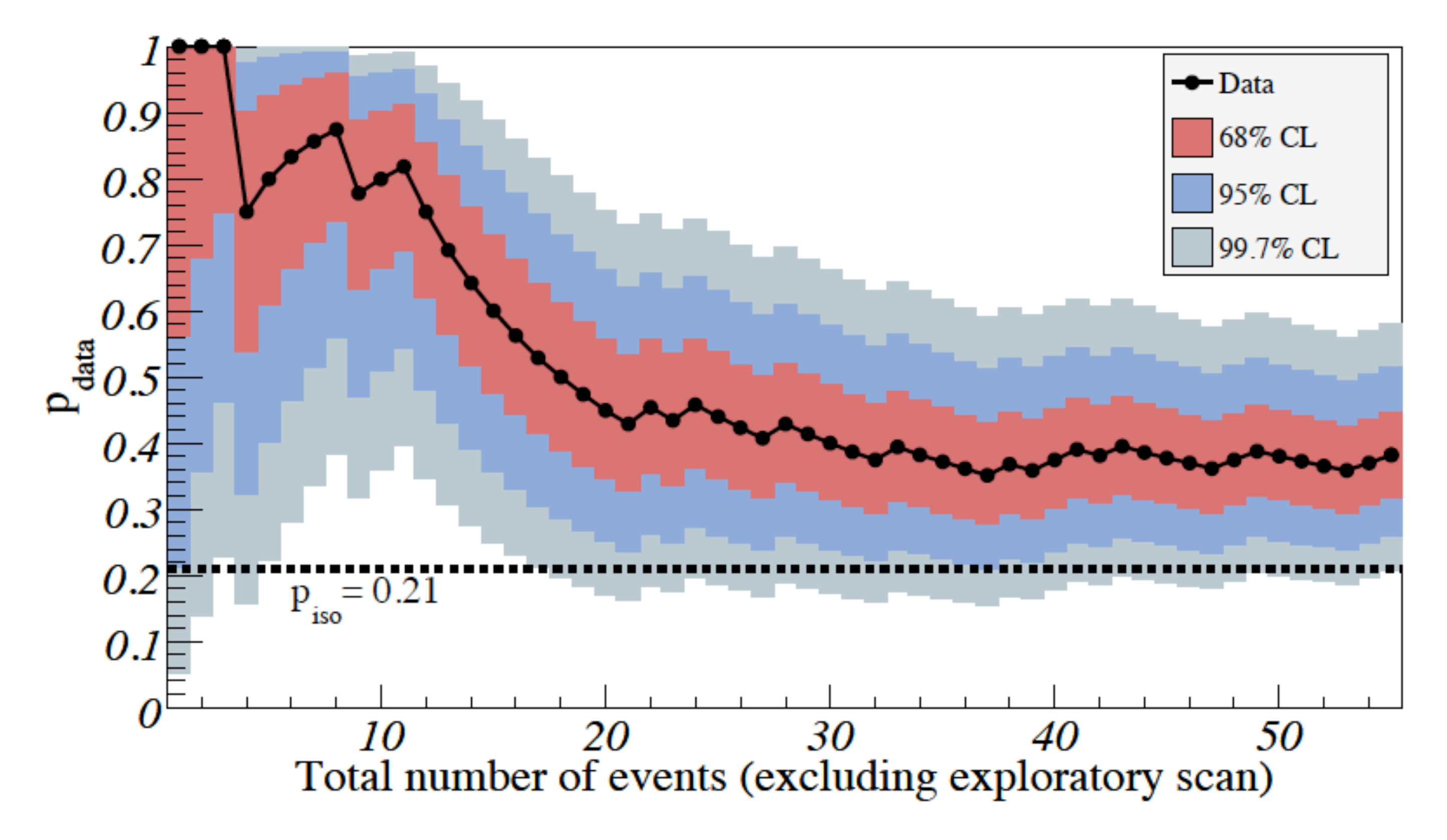}
\caption{Left: The 69 arrival directions of cosmic rays with energy $E>55$~EeV detected by the Pierre Auger Observatory up to December 2009 are plotted as black dots in an Aitoff-Hammer projection of the sky in galactic coordinates. The solid line represents the field of view of the Southern Observatory for zenith angles smaller than $60^\circ$. Blue
circles of radius $3.1^\circ$ are centred at the positions of the 318 AGN in the VCV catalogue that lie within 75~Mpc
and that are within the field of view of the Observatory. Darker blue indicates larger relative exposure. The
exposure-weighted fraction of the sky covered by the blue circles is 21\%. Right: Fraction of events correlating with AGN as a function of the cumulative
number of events, starting after the exploratory data. The expected correlating fraction
for isotropic cosmic rays is shown by the dotted line. From \Ref~\cite{Abraham2010b}}
\label{anisotropia}
\end{figure}

The most recent discussion of anisotropies in the sky distribution of ultrahigh energy
events began when the Pierre Auger Observatory reported a correlation of its highest energy events with AGN~\cite{Abraham2007} in the 12th Veron-Cetty \& Veron (VCV) catalogue~\cite{VCV}. To calculate a meaningful statistical significance in such an analysis, it is important to define the search procedure {\it a priori} in order to ensure it is not inadvertently devised especially to suit the particular data set after having studied it. With the aim of avoiding accidental bias on the number of trials performed in selecting the cuts, the Auger anisotropy analysis scheme followed a pre-defined process. First an exploratory data sample was employed for comparison with various source catalogues and for tests of various cut choices.  The results of this exploratory period were then used to design prescriptions to be applied to subsequently gathered data. The first 14 events were used for an
exploratory scan and the correlation was most
significant for AGN for energy threshold $5.5 \times10^{19}$eV with redshifts $z< 0.018$ (distances $<75$ Mpc) and within $3.1^\circ$  separation angles.  The subsequent 13 events established a 99\% confidence level
for rejecting the hypothesis of isotropic cosmic ray flux. The reported fraction of correlation events was $69^{+11}_{-13}\%$. An analysis with data
up to the end of 2009 (69 events in total, as seen in the left panel of Figure~\ref{anisotropia}) indicated that the correlation level decreased to
$38^{+7}_{-6}\%$~\cite{Abraham2010b}.  In the right panel of Figure~\ref{anisotropia} we show the most likely value of the fraction of the correlated events with objects in the VCV catalogue as a function  of the total number of time-ordered events (the
events used in the exploratory scan are excluded). The $1\sigma$ and $2\sigma$ uncertainties in this value are indicated. The current estimate
of the fraction of correlating cosmic rays is $33 \pm 5\%$ (28 events
correlating from a total of 84 events)  with 21\% expected
under the isotropic hypothesis~\cite{Kampert2011}.  

The Telescope Array Collaboration has also searched for correlation with AGN in the VCV catalogue~\cite{AbuZayyad:2012hv, Fukushimaicrc}.  The TA exposure is peaked in the Northern hemisphere
so the AGN visible to TA are not the same as the ones visible to Auger, though there is some overlap.
When the distribution of nearby AGN is taken into account, and assuming equal AGN luminosities in UHECR, the correlating fraction would be 40\%.

A complete report on the current status for anisotropy searches can be found in~\cite{Deligny:2013xta}. The report includes,  in the region around $10^{18}$eV, constraints from measuring
the first harmonic modulation in the right ascension distribution of arrival directions, and search for point-like sources that would be indicative of a flux of neutrons (see also \Ref~\cite{AU1}); at higher energies, searches for clustering in arrival directions, and 
correlations with nearby extragalactic objects  (see also \Ref~\cite{AU2}) or the large scale
structure of the Universe.

\section{Mass composition estimate: the biggest challenge}
\label{mass}

A determination of primary composition is invaluable in revealing the origin of 
cosmic rays as this information would provide important bounds on sources and on possible production 
and acceleration mechanisms. 
In addition, a proper interpretation of anisotropy information requires knowledge of the primary mass 
due to the influence on propagation of the galactic and intergalactic magnetic fields.
A detailed analysis of composition data from various experiments has been presented
in \Ref~\cite{Kampert:2012mx}. We first present a brief description of the 
general signatures of the EAS (See \Ref~\cite{Anchordoqui:2004xb} for a summary of 
the phenomenology of these giant air showers). After that, we introduce the shower observables sensitive to primary species.
\subsection{Signatures of Extensive Air Showers}

The evolution of an extensive air shower is dominated by electromagnetic
processes. The interaction of a 
baryonic cosmic ray  with an air nucleus high in the atmosphere leads to a 
cascade of secondary mesons and nucleons. The first few 
generations of charged pions interact again, producing a hadronic core, 
which continues to feed the electromagnetic and muonic components of the 
showers. Up to about $50$~km above sea level,
the density of atmospheric target nucleons is $n \sim 
10^{20}$~cm$^{-3},$ and so even for relatively
low energies, say  $E_{\pi^{\pm}}\approx 1$~TeV, the probability 
of decay before interaction falls below 10\%. 
Ultimately, the electromagnetic cascade dissipates around 90\%
of the primary particle's energy, and hence the total number of 
electromagnetic particles is very nearly proportional to the shower energy.

By the time a vertically incident 
$10^{20}$eV proton shower
reaches the ground, there are about $10^{11}$ secondaries with energy above
90~keV in the the annular region extending 8~m to 8~km from the shower core.
Of these, 99\%  are photons, electrons, and positrons, with a typical ratio 
of $\gamma$ to $e^+  e^-$ of 9 to 1. Their mean energy  is 
around 10~MeV and they transport 85\% of the total energy at ground level. Of course, 
photon-induced showers are even more dominated by the electromagnetic channel, 
as the only significant muon generation mechanism in this case is the decay of
charged pions and kaons produced in $\gamma$-air  interactions~\cite{Mccomb:tp}. 

It is worth mentioning that these figures dramatically change for the case of 
very inclined showers. For a primary zenith angle, $\theta > 70^{\circ},$ the 
electromagnetic component becomes attenuated exponentially with atmospheric 
depth, being almost completely absorbed at ground level. 
As a result, most of the energy at ground level from an inclined shower is
carried by muons.

In contrast to  hadronic collisions, the electromagnetic 
interactions of shower particles can be calculated very accurately from 
quantum electrodynamics. Electromagnetic interactions are thus not a 
major source of systematic errors in shower simulations. The first 
comprehensive treatment of electromagnetic showers was elaborated by 
Rossi and Greissen~\cite{Rossi}.  This treatment was recently cast in a more 
pedagogical from by Gaisser~\cite{Gaisser:vg}, which we summarize in the 
subsequent paragraphs.

The generation of the electromagnetic component is 
driven by electron bremsstrahlung and pair production~\cite{Bethe:1934za}.
Eventually the average energy per particle drops below a critical energy,
$\epsilon_0$, at which point ionization takes over from bremsstrahlung and pair
production as the dominant energy loss mechanism. The $e^\pm$ energy loss rate due to
bremsstrahlung radiation is nearly proportional to their energy, whereas the 
ionization loss rate varies only logarithmically with the $e^\pm$ energy.
Throughout this note we take the critical energy 
to be that at which the ionization loss per radiation length
is equal to the electron energy, yielding $\epsilon_{0} = 710~{\rm MeV}/(Z_{\rm eff} +0.92) \sim 
$~86~MeV~\cite{Rossi:book}. The changeover 
from radiation losses to ionization losses depopulates the shower.
One can thus categorize the shower development in three phases: the growth phase, in which all the particles 
have energy $> \epsilon_0$; the shower maximum, $X_{\rm max}$; and the shower 
tail, where the particles only lose energy, get absorbed or decay. 

Most of the general features of an electromagnetic cascade can be understood in terms of the toy model 
due to Heitler~\cite{Heitler}. In this model, the shower is imagined to develop exclusively 
via bremsstrahlung and pair production, each of which results in the conversion of one particle into two.
These physical processes are characterized by an 
interaction length $X_0$. One can thus imagine the shower as a particle tree with branches that
bifurcate every $X_0$, until they fall below a critical energy, $\epsilon_0$, at which point
energy loss processes dominate. 
Up to $\epsilon_0$, the number of particles grows geometrically, so that after 
$n = X/X_0$ branchings, the total number of particles in the shower is 
$N \approx 2^n$.  At the depth of shower maximum $X_{\rm max}$, all particles are at 
the critical energy, $\epsilon_0$, and the energy of the primary particle, $E_0$, is 
split among all the $N_{\rm max} = E_0 / \epsilon_0$ particles.
Putting this together, we get:
\begin{equation} \label{heitler}
X_{\rm max} \approx X_0 \,\, \frac{\ln(E_0/\epsilon_0)}{\ln 2} \,\,.
\end{equation}

Even baryon-induced showers are dominated by electromagnetic processes, so 
this toy model is still enlightening for such cases.  In particular, for proton 
showers, Eq.~(\ref{heitler}) tells us that the $X_{\rm max}$ scales logarithmically
with primary energy, while $N_{\rm max}$ scales linearly. Moreover, to extend 
this discussion to heavy nuclei, we can apply the superposition principle as 
a reasonable first approximation. In this approximation, we pretend that the nucleus
comprises unbound nucleons, such that the point of first interaction of one nucleon is
independent of all the others.  Specifically, a shower produced by a nucleus with energy 
$E_{_A}$ and mass $A$ is modelled by a collection of $A$ proton showers, each with $A^{-1}$ of the
nucleus energy. Modifying Eq.~(\ref{heitler}) accordingly one easily obtains
$X_{\rm max} \propto \ln (E_0/A)$.

Changes in the mean mass composition of the cosmic ray flux as a function of energy will 
manifest as changes in the mean values of $X_{\rm max}$. This change of $X_{\rm max}$ 
with energy\footnote{The elongation rate is commonly reported per decade of energy,
$D_{10} = \partial \langle X_{\rm max} \rangle/\partial \log E$, where $D_{10} = 2.3 D_e.$ }
is commonly known as the elongation rate theorem~\cite{Linsley:P3}:
\begin{equation} 
D_e = \frac{\delta X_{\rm max}}{\delta \ln E} \,\,.\end{equation}
For purely electromagnetic showers, $X_{\rm max}(E) \approx  X_0\, \ln(E/\epsilon_0)$ and
then the elongation rate is $D_e \approx X_0$. For proton primaries, the multiplicity rises 
with energy, and thus
the resulting elongation rate becomes smaller.  This can be understood by noting that, on average,
the first interaction is determined by the proton mean free path in the atmosphere, $\lambda_N$.
In this first interaction the incoming proton splits into $\langle n(E) \rangle$ secondary particles, 
each carrying an average energy $E/\langle n(E) \rangle$. Assuming that $X_{\rm max}(E)$ 
depends logarithmically on energy,
as we found with the Heitler model described above, it follows that, 
\begin{equation}
X_{\rm max}(E)=\lambda_N + X_0\, \ln[E/\langle n(E) \rangle]\,\, .
\end{equation}
If we assume a multiplicity dependence $\langle n(E) \rangle \approx n_0 E^{\Delta}$, then  
the elongation rate becomes,
\begin{equation} 
\frac{\delta X_{\rm max}}{\delta \ln E}= X_0\,\left[1-\frac{\delta \ln \langle n(E) 
\rangle}{\delta \ln E} \right] + \frac{\delta \lambda_{N}}{\delta \ln E}
\end{equation}
which corresponds to the form given in~\cite{Linsley:gh},
\begin{equation} 
D_e = X_0 \,\left[ 1-\frac{\delta \ln \langle n(E) \rangle}{\delta \ln E} + 
\frac{\lambda_{N}}{X_0} \frac{\delta \ln(\lambda_{N})}{\delta \ln E} \right] =  X_0\,(1-B) \,\,.
\label{ERLW}
\end{equation}
Using the superposition model and assuming that 
\begin{equation}
B \equiv \Delta - \frac{\lambda_{N}}{X_0} \,\,\frac{\delta \ln \lambda_{N}}{\delta \ln E}
\label{theB}
\end{equation} 
is not changing with energy, one obtains for mixed primary 
composition~\cite{Linsley:gh}
\begin{equation}  
D_e =\, X_0\,(1-B)\,
\left[1 - \frac{\partial \langle \ln A \rangle }{\partial \ln E} \right]\, .
\label{er}
\end{equation}
Thus, the elongation rate provides a measurement of the change of the 
mean logarithmic mass with energy.   

In \Ref~\cite{Matthews:2005sd}, a precise calculation of a hadronic shower evolution has been presented assuming  that hadronic interactions produce exclusively pions. 
The first interaction diverts 1/3 of the available energy ($E_0/3$) into the EM component via the $\pi^0$'s, while the remaining 2/3 continue as hadrons.  Using $pp$ data~\cite{Amsler:2008zzb},
we parametrized the charged particle production in the first interaction as
$N_{\pi^{\pm}} = 41.2(E_0/1~{\rm PeV})^{1/5}$. The depth of shower maximum is thus the same as for an electromagnetic shower of energy $E_0/(3 N_{\pi^\pm})$, giving for a proton initiated shower:
\begin{eqnarray}
X_{\rm max}^p  & = & X_0 + X_{_{\rm EM}}  \, \ln[E_0/( 6 N_\pi \epsilon_0)]  \nonumber \\
                          & = & (470 + 58 \, {\rm log_{10}} [E_0/1~{\rm PeV}])~{\rm g/cm}^2 \,.
\label{Xmax-Matthews2}
\end{eqnarray}
For protons the elongation rate results $\approx 58 \,{\rm g/cm}^2 $ per decade of energy, in good agreement with calculations that model the shower development using the best estimates of the relevant features of the hadronic interactions. Muons are produced from the pion decay when they reach the critical energy (  $\xi^\pi_{\rm c}$) after $n_c $ generations.  Introducing $\beta=\ln(2N_\pi)/\ln(3N_\pi)$, the total number of muons is:
\begin{equation}\label{hs} 
N_\mu = (E_0/\xi^\pi_{\rm c})^\beta \, . 
\end{equation} 
For $N_\pi=5$, $\beta = 0.85$. Unlike the electron number, the muon multiplicity does not grow linearly with the primary energy, but at a slower rate.
The precise value of $\beta$ depends on the average pion multiplicity used.  It also depends on the inelasticity of the hadronic interactions. The critical pion energy  $\xi^\pi_{\rm c}\approx 20 $GeV in a shower generated by 1 PeV proton.

Using the superposition model, we obtain for a nucleus of mass A. 
\begin{equation}
N_{\mu}^{A} = A\, \left[\frac{(E_0/A)}{\xi^\pi_{\rm c}}\right]^\beta \, . 
\label{hsA} 
\end{equation} 

From the discussion above, it follows that the depth of shower maximum and the number of muons depend on the mass of the primary particle: iron initiated showers develop faster in the atmosphere, having smaller $X_{\rm max} $ than proton initiated shower, while larger number of muons are expected for heavier nuclei.

While the  Heitler model is very useful for imparting a first intuition regarding global shower properties, 
the details of shower evolution are far too complex to be fully described
by a simple analytical model. Full Monte Carlo simulation of interaction and transport of each individual
particle is required for precise modelling of the shower development. At present two Monte Carlo  
packages are available to simulate EAS: {\sc corsika} (COsmic Ray SImulation for 
KAscade)~\cite{Heck:1998vt} and {\sc aires} (AIR shower Extended Simulation)~\cite{Sciutto:1999jh}.  
Both programs provide fully 4-dimensional simulations of the air showers initiated by protons, photons, 
and nuclei. A comparative study using these codes can be found in \Ref~\cite{Knapp:2002vs}. Different hadronic interaction models are used in these event generators,  such as {\sc sibyll}~\cite{Fletcher:1994bd}, 
{\sc qgsjet}~\cite{Kalmykov:te}  and {\sc epos}~\cite{epos1, epos2}. The LHC data, particularly those measured in the extreme
forward region of the collisions, is of great importance to the physics of EAS. As an example,  {\sc epos} has been modified to reproduce in
detail LHC data from various experiments~\cite{eposLHC}.

\subsection{Measurement of mass sensitive observables}

In this section, we discuss how baryonic species may, to some extent, be distinguished by the
signatures they produce in the atmosphere.  The estimate of primary masses is the
most challenging task in high energy cosmic ray physics as such measurements rely on comparisons of
data to models.  EAS simulations are subject to uncertainties mostly because hadronic interaction models need to be extrapolated at energy ranges several order of magnitude higher than those accessible to current particle accelerators.  In what follows, we consider both surface array
and fluorescence detector observables.

\begin{figure}[t]
\begin{center}
\includegraphics[width=0.78\textwidth]{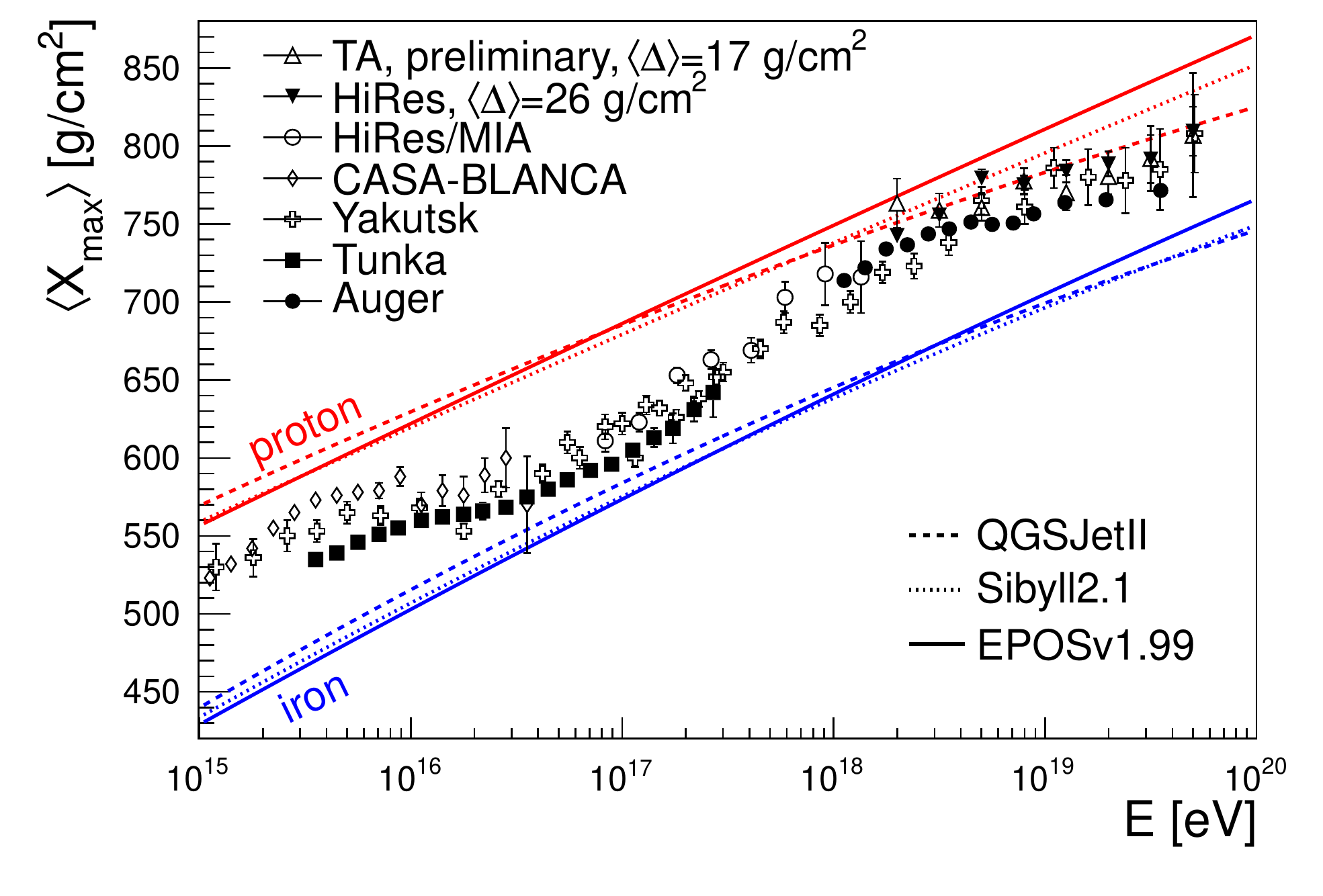}
\end{center}
\caption{Measurements of $\langle X_{\rm max}\rangle$ with non-imaging Cherenkov detectors (Tunka~\cite{Tunka}, Yakutsk~\cite{Yakutsk}, CASA-BLANCA~\cite{CASA}) and fluorescence detectors (HiRes/MIA~\cite{HiResMIA}, HiRes~\cite{HiRes}, Auger~\cite{Facal} and TA~\cite{TAER} compared to air shower simulations using hadronic interaction models. HiRes and TA data have been corrected for detector effects as indicated by the $\langle \Delta \rangle$  values, to allow comparison with the unbiased measurement from Auger.  This picture is taken from \Ref~\cite{Kampert:2012mx}. }
\label{elong}
\end{figure}

The main purpose of fluorescence detectors is to measure the properties of the longitudinal development. The shower longitudinal profile is usually parameterized with a function, such as the Gaisser-Hillas 
function~\cite{Gaisser:icrc} used by the Pierre Auger Observatory. Using this parametrization, fluorescence detectors can measure
$X_{\rm max}$ with a statistical precision typically around $30~{\rm g} / {\rm cm}^2$.  The speed of shower development is the clearest indicator of the primary composition. It was 
shown in Sec.~\ref{mass} using the superposition model that  there is a difference between the 
depth of maximum in proton and iron induced showers. In fact, nucleus-induced showers develop faster, 
having $X_{\rm max}$ higher in the atmosphere. From Monte Carlo simulations, one finds
that the difference between the 
average $X_{\rm max}$ for protons and iron nuclei is about 90 -- 100~g/cm$^2$. However, 
because of shower-to-shower fluctuations, it is not possible to obtain meaningful 
composition estimates from $X_{\rm max}$ on a shower-by-shower basis, though one can derive composition
information from the magnitude of the fluctuations themselves.    
For protons, the depth of first interaction fluctuates more than it does for iron, and 
consequently the fluctuations of $X_{\rm max}$ are larger for protons as well. 
In Figure~\ref{elong} the $ \langle X_{\rm max} \rangle$ 
measurements of $\langle X_{\rm max}\rangle$ with non-imaging Cherenkov detectors (Tunka~\cite{Tunka}, Yakutsk ~\cite{Yakutsk}, CASA-BLANCA~\cite{CASA}) and fluorescence detectors (HiRes/MIA~\cite{HiResMIA}, HiRes~\cite{HiRes}, Auger~\cite{Facal} and TA~\cite{TAER} compared to air shower simulations using several hadronic interaction models are presented.  The conclusion of the detailed study in \Ref~\cite{Kampert:2012mx} indicates that, around the region of the ankle of the cosmic
ray spectrum, the measurements are compatible within their
quoted systematic uncertainties and the $\langle X_{\rm max}\rangle$ is close to the prediction for air showers initiated by a predominantly light
composition.  However, at higher energies, the experimental uncertainties
are still too large to draw conclusions from the data. In addition, the
systematic differences between different type of measurements are very sensitive to the
particular interaction model used for the interpretation. 

The electromagnetic component of an EAS suffers more scattering and energy loss than the muonic component and consequently, muons tend to arrive earlier and over a shorter period of time. This means that parameters characterizing the time structure of the EAS, as measured by surface arrays,  
will be correlated with $X_{\rm max}$ and hence with primary mass.  An early study of the shower signal 
observed in water \v{C}erenkov detectors 
arrays~\cite{Watson:ja} established the utility of a 
shower property known as risetime in estimating the primary composition. 
Specifically, the risetime, $t_{1/2}$, is defined as the time for the 
signal to rise from 10\% to 50\% of the full signal. 

\begin{figure}[t]
\begin{center}
\includegraphics[width=0.6\textwidth]{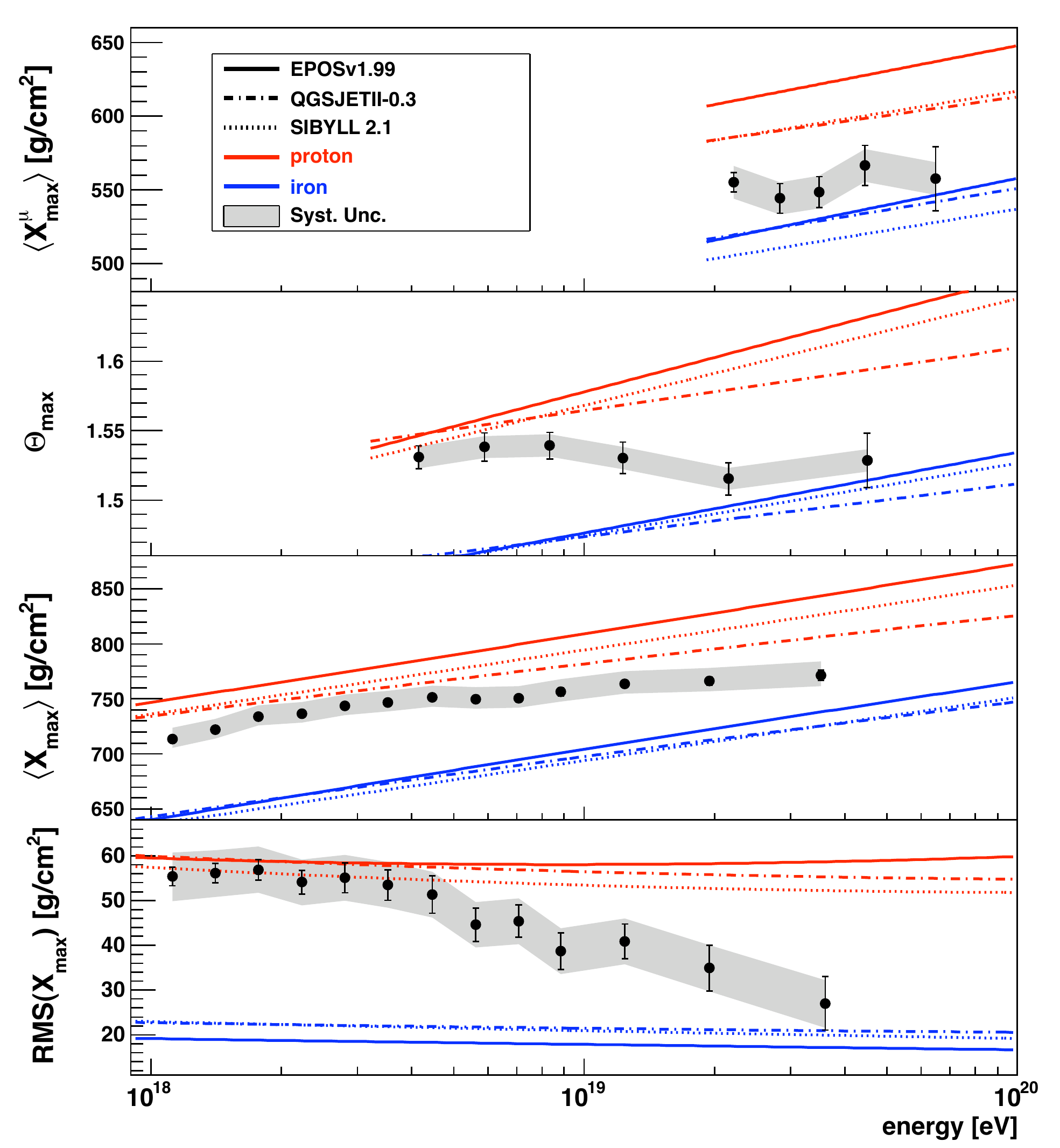}
\end{center}
\caption{From top to bottom,$\langle X_{\rm max}^\mu
          \rangle$, $\Theta_{\rm{max}}$, $\langle X_{\rm max}\rangle$ and RMS ($\langle X_{\rm max}\rangle$) as a function of Energy compared with air
          shower simulations using different hadronic
          interaction models. The error bars
          correspond to the statistical uncertainty, the grey areas correspond to the systematic
          uncertainty~\cite{Diego:2011}. Updated studies of $X_{\rm max}^\mu$, $ \langle X_{\rm max} \rangle$ and RMS($X_{\rm max}$) can be found in \Ref~\cite{Auger:icrc2013}.  }
\label{elong2}
\end{figure}

In ground array experiments the analysis is usually performed by projecting the signals 
registered by the detectors into the shower plane (perpendicular to the shower axis)
and thus, neglecting the further shower evolution of the late regions. 
As a consequence, for inclined showers, the circular symmetry in the signals of surface detectors 
is broken. This results in a dependence of the signal 
features on the azimuth angle in the shower plane~\cite{Dova:2003rz, Dova:2001jy}. A detailed study based on Monte
Carlo simulations~\cite{Dova:2009fk}, showed that for showers
arriving with zenith angle $\theta >30^{\circ}$, this is mainly due to
the attenuation of the electromagnetic  component of the shower as
it crosses additional atmosphere to reach a late detector. 
For a given primary energy $E$, the risetime asymmetry in water \v{C}erenkov detectors 
array, as in the Pierre Auger Observatory,  depends on zenith angle $\theta$
of the primary cosmic ray in such a way that its behaviour versus $\sec \theta$  
is reminiscent of the longitudinal development of the shower.  In \Ref~\cite{Dova:2009fk}, it was shown that the zenith angle at which the risetime asymmetry becomes maximum, $\Theta_{max}$, is correlated with the shower development and hence with the primary species.

Using the time information of the signals recorded by the water \v{C}erenkov detectors, it is also possible to obtain information about the longitudinal development of the hadronic component of extensive air showers and the first interaction point in an indirect way. In particular, a method was developed to reconstruct the Muon Production Depth (MPD), the distance to the production of the muon measured parallel to the shower axis, using the signals of detectors far from the core~\cite{cazon_2004}. The MPD technique allows one to convert the time distribution of the signal recorded by the SD detectors into muon production distances using an approximate relation between production distance, transverse distance and time delay with respect the shower front plane. From the MPDs a new observable can be defined, $X_{\rm max}^\mu$, as the depth along the shower axis where the number of produced muons reaches a maximum, which is sensitive to primary mass.

The evolution of $X_{\rm max}^\mu$, $\Theta_{max}$, $ \langle X_{\rm max} \rangle$ and RMS($X_{\rm max}$) with energy, as measured by the Pierre Auger Observatory with data up to 2010~\cite{Diego:2011}, is presented in Figure~\ref{elong2}. For a very complete discussion of these results see \Ref~\cite{Unger:2013qya}. It is worth noting that the  these analyses come from completely independent techniques that have different sources of systematic uncertainties.  Concerning the RMS, a variety of compositions can give rise to large values
of the RMS, because the width of the $X_{\rm max}$ is influenced by both,
the shower-to-shower fluctuations of individual components and their relative
displacement in terms of $\langle X_{\rm max}\rangle$.  These measurements from Auger may be interpreted as a transition
to a heavier composition that may be caused by a Peters-cycle~\cite{Peters}
in extragalactic sources similar to what has been observed at
around the knee~\cite{Kampert:2012mx,Unger:2013qya}.  

Updated studies of $X_{\rm max}^\mu$, $ \langle X_{\rm max} \rangle$ and RMS($X_{\rm max}$) from the Pierre Auger Observatory can be found in \Ref~\cite{Auger:icrc2013}. The most recent results on $ \langle X_{\rm max} \rangle$ measurements from the TA experiment were presented in \Refs~\cite{TA1:icrc2013, TA2:icrc2013}.


\section*{Acknowledgments} I would like to thank the organizers of the 2013 CERN-Latin-American School of HEP for the excellent and stimulating school. I am indebted to Jim Cronin who introduced me to the fascinating world of cosmic rays, he has been an inspiration to me. I also would like to thank Luis Anchordoqui,  Luis Epele and John Swain for the many years of fruitful discussions on the phenomenology of EAS and propagation of UHECR {\it en route} to us from their sources. I am grateful to Hernan Wahlberg, Paul Sommers, Michael Unger, Alan Watson, Analisa Mariazzi, Diego Garc{\'\i}a-Pinto, Fernando Arqueros, Tom Paul and all my colleagues from the Pierre Auger Observatory for   lively and enlightening discussions.

\end{document}